\documentclass[12pt]{article}

\usepackage{graphicx}

\usepackage{natbib}
\usepackage{ametsoc}
 
\renewcommand{\bm}[1]{{\bf #1}}
 
\begin{document}
\title{Oceanic Internal Wave Field: Theory of Scale-invariant Spectra}
\author{
\centerline{\sc{Yuri V. Lvov}}\\
\centerline{\sc{Rensselaer Polytechnic Institute, Troy, NY USA}}
%
\and
\centerline{\sc{Kurt L. Polzin}}\\
\centerline{\sc{Woods Hole Oceanographic Institution, Woods Hole, MA USA}}
\and
\centerline{\sc{Esteban G. Tabak}}\\
\centerline{\sc{New York University, New York, NY USA}}
\and
\centerline{\sc{Naoto Yokoyama} \thanks{\textit{Corresponding author present address:}
Department of Aeronautics and Astronautics,
Graduate School of Engineering,
Kyoto University,
Kyoto 606-8501 JAPAN
\newline{E-mail: yokoyama@kuaero.kyoto-u.ac.jp}}}\\
\centerline{\sc{Doshisha University, Kyotanabe, Kyoto Japan}}
}
\amstitle



\begin{abstract}
Steady scale-invariant solutions of a kinetic equation
describing the statistics of oceanic internal gravity waves 
based on wave turbulence theory are investigated.
It is shown in the non-rotating scale-invariant limit that the
collision integral in the kinetic equation
diverges for almost all spectral power-law exponents.
These divergences come from resonant
interactions with the smallest horizontal  wavenumbers and/or the largest horizontal wavenumbers
with extreme scale-separations.

We identify a small domain in which the scale-invariant collision
integral converges and numerically find
a convergent power-law solution.  This numerical solution is close to the
Garrett--Munk spectrum.
Power-law exponents which potentially permit 
a balance between the infra-red and ultra-violet 
divergences are investigated. 
The balanced exponents are generalizations of an exact solution of the 
scale-invariant 
kinetic equation, the Pelinovsky--Raevsky spectrum. 

A balance between oppositely signed divergences states that infinity minus infinity may be approximately equal to zero. 
A small but finite Coriolis parameter  representing the effects of rotation is introduced into the kinetic equation to determine solutions over the divergent part of the domain using rigorous asymptotic arguments.  This gives rise to the induced diffusion regime.

The derivation of the kinetic equation is based on an assumption of weak nonlinearity. 
Dominance of the nonlocal interactions puts the self-consistency
of the kinetic equation at risk.  Yet these weakly nonlinear stationary states are consistent with much of the observational evidence. 
\end{abstract}

\section{Introduction}

Wave-wave interactions in continuously stratified fluids have been a
subject of intensive research in the last few decades. Of particular
importance is the observation of a nearly universal internal-wave energy spectrum
in the ocean, first described by Garrett and Munk~\citep{GM72,GM75,cairns1976iwo,GM_ARF}.     However, it appears that ocean is too complex to be described by one
universal model.  Accumulating evidence suggests that there is
measurable variability of observed experimental spectra~\citep{iwobsPL}.
In particular, we have analyzed the decades of observatioanl programs, and 
we have come to a conclusions that the high-frequency--high-wavenumber part of the spectrum,
can be characterized through simple power law fits with variable
exponents.

It is generally thought~\citep{muller1986nia,O74,olbers1976net,iwobsPL} that nonlinear interactions significantly contribute to determining the background oceanic spectrum and this belief motivates the investigation of spectral evolution equations for steady balances.  A particularly important study in this regard is the demonstration that the Garrett and Munk vertical wavenumber spectrum is stationary and supports a constant downscale energy flux 
\citep{mccomas-1981} associated with resonant interactions.  The accumulating evidence alluded to above suggests there is more to the story.  

The purpose of the present study is to lay down a firm theoretical
framework that allows a detailed analysis of power-law spectra of
internal waves in the ocean. We investigate the parameter space of the possible power laws 
with a specific focus upon extreme scale separated interactions and their role in dominating spectral transfers.  We then use this theoretical framework to interpret the observed oceanic variability.  

Due to the quadratic nonlinearity
of the underlying fluid equations and dispersion relation allowing three-wave resonances, 
internal waves interact through triads.
In the weakly nonlinear regime, the nonlinear
interactions among internal waves concentrate on their {\em resonant\/} set, and 
can be described by a kinetic equation, which assumes the
familiar form~\citep{caillol2000kea,H66,K66,K68,gm_lvov,lvov-2004-195,mccomas-1977-82,milder,MO75,O74,olbers1976net,pelinovsky1977wti,pomphrey1980dni,voronovich1979hfi,zak_book}:

\begin{eqnarray}
\frac{\partial n_{\bm{p}}}{\partial t} \! &=& \!
 4\pi \int d \bm{p}_{12}
\left(
 |V_{\bm{p}_1,\bm{p}_2}^{\bm{p}}|^2 \, f^{\bm{p}}_{\bm{p}_1\bm{p}_2} \,
\delta_{{\bm{p} - \bm{p}_1-\bm{p}_2}} \, \delta_{\omega_{\bm{p}}
-\omega_{{\bm{p}_1}}-\omega_{{\bm{p}_2}}}
\right.
\nonumber \\
&&
\left.
- |V_{\bm{p}_2,\bm{p}}^{\bm{p}_1}|^2\, f^{\bm{p}_1}_{\bm{p}_2\bm{p}} \, \delta_{{\bm{p}_1 - \bm{p}_2-\bm{p}}} \,
  \delta_{{\omega_{\bm{p}_1} -\omega_{\bm{p}_2}-\omega_{\bm{p}}}}
-
 \, |V_{\bm{p},\bm{p}_1}^{\bm{p}_2}|^2\, f^{\bm{p}_2}_{\bm{p}\bm{p}_1}\, \delta_{{\bm{p}_2 - \bm{p}-\bm{p}_1}} \,
  \delta_{{\omega_{\bm{p}_2} -\omega_{\bm{p}}-\omega_{\bm{p}_1}}}
\right)
 \, ,\nonumber\\
&&
\mathrm{with} \quad f^{\bm{p}}_{\bm{p}_1\bm{p}_2} = n_{\bm{p}_1}n_{\bm{p}_2} -
n_{\bm{p}}(n_{\bm{p}_1}+n_{\bm{p}_2}) \, .
\label{KineticEquation}
\end{eqnarray}
Here $n_{\bm{p}} = n(\bm{p})$ is a three-dimensional action spectrum (see Eq.~(\ref{WaveAction})) with wavenumber
$\bm{p} = (\bm{k}, m),$
i.e.\ $\bm{k}$ and $m$ are the horizontal and vertical components of $\bm{p}$.
Action or wave-action can be viewed as ``number'' of waves with a given
wavenumber.
The frequency $\omega_{\bm{p}} $ is given by a linear dispersion relation 
(\ref{InternalWavesDispersion}) below. 
Consequently, wave action multiplied by frequency $\omega_{\bm{p}} n_{\bm{p}}$
can be seen as quadratic spectral energy density of internal waves. Note that 
the wavenumbers are signed variables, while
the wave frequencies are always positive.  The factor
$V^{\bm{p}}_{\bm{p}_1\bm{p}_2}$ is the interaction matrix element
describing the transfer of wave action among the members of a
triad composed of three wave vectors ${\bm{p}}={\bm{p}_1}+ {\bm{p}_2}$.

Following Kolmogorov's viewpoint of energy cascades
in isotropic Navier--Stokes turbulence,
one may look for statistically stationary states using scale-invariant solutions to the 
kinetic equation~(\ref{KineticEquation}). The solution may occur in an inertial
subrange of wavenumbers and frequencies that are far from those where forcing and
dissipation act, and also far
from characteristic scales of the system, including  
the Coriolis frequency 
resulting from the rotation of the Earth, the buoyancy frequency due
to stratification and the ocean depth.
Under these assumptions, the dispersion relation and the interaction matrix elements are 
locally scale-invariant.  It is natural, therefore, in this restricted domain, to
look for self-similar solutions of Eq.~(\ref{KineticEquation}), which 
take the form
\begin{eqnarray}
   n(\bm{k}, m)=  |\bm{k}|^{-a}  |m|^{-b}.
\label{PowerLawSpectrum}
\end{eqnarray}
Values of $a$
and $b$ for which the right-hand side of
Eq.~(\ref{KineticEquation}) vanishes identically
correspond to steady solutions of the kinetic
equation, and hopefully also to statistically steady states of the ocean's
wave field. 
Unlike Kolmogorov turbulence, 
the exponents which give steady solutions
can not be determined by the dimensional analysis alone (see, for example, \cite{polzin-2004}). This is the case 
owing to multiple characteristic length scales in anisotropic systems. 

Before seeking steady solutions, however,
one should find out whether the
improper integrals~\footnote{Improper integrals have the form
\begin{eqnarray} 
\lim\limits_{b\to\infty}\int\limits_a^b f(x) d x, {\ \  \ }
\lim\limits_{a\to-\infty}\int\limits_a^b f(x) d x, {\ \  \ }
\lim\limits_{c\to b^-}  \int\limits_a^c f(x) d x, {\ {\rm or}  \ } 
\lim\limits_{c\to a^+}  \int\limits_c^b f(x) d x. {\ \  \ } 
\nonumber\end{eqnarray}
When the limit exists (and is a number), the improper integral is called 
{\it convergent}; when  the limit does not exist or it is infinite, the improper integral is called {\it divergent}.
}
in the kinetic equation~(\ref{KineticEquation}) converge. 
This is related to the question of {\em locality\/} of the
interactions: a convergent integral characterizes the physical scenario where 
interactions of neighboring wavenumbers dominate the evolution of the wave spectrum, 
while a divergent one 
implies that distant, nonlocal interactions in the wavenumber space dominate.

In the present manuscript we demonstrate analytically
that the internal-wave collision integral diverges
for {\em almost\/} all values of $a$ and $b$. In particular, the collision integral has an 
infra-red (IR) divergence 
at zero, i.e. $|\bm{k}_1|$ or $|\bm{k}_2|$ $\to 0$ and an ultra-violet (UV) divergence
at infinity, i.e. $|\bm{k}_1|$ and $|\bm{k}_2|$ $\to \infty$.
Thus IR divergence comes from interactions with the smallest
wavenumber, and UV divergence comes from interactions with the largest
horizontal wavenumber.
There is only one exception where the integral converges:
the segment with $b=0$ and  $7/2 < a < 4$.
The $b=0$ line corresponds to wave action independent of vertical wavenumbers, 
$\partial n / \partial m = 0$.
Within this segment we numerically determine a new 
steady {\em convergent\/} solution to
Eq.~(\ref{KineticEquation}), with
\begin{eqnarray} n(\bm{k},m) \propto  |\bm{k}|^{-3.7}.
\label{NewPoint}
\end{eqnarray}
This solution is not far from the large-wavenumber form of the Garrett--Munk (GM) spectrum~\citep{GM72,GM75,cairns1976iwo,GM_ARF}: 
\begin{equation}
n(\bm{k},m) \propto  |\bm{k}|^{-4}~.
\label{GM}
\end{equation}

Alternatively, one 
can explore
the physical interpretation of divergent solutions. 
We find a region in $(a,b)$ 
space where there are both IR and UV divergences having
opposite signs.
This suggests a possible scenario where the
two divergent contributions may cancel each other, yielding a steady
state. An example of such a case 
is provided by the Pelinovsky--Raevsky (PR) spectrum~\citep{pelinovsky1977wti},
\begin{eqnarray}
 n_{\bm{k},m} \propto |\bm{k}|^{-7/2}
 |m|^{-1/2}
.
\label{LT}
\end{eqnarray}

This solution, however, is only one among infinitely many. 
The problem at hand is a generalization of the
concept of principal value integrals:
for $a$ and $b$ which give opposite signs of the divergences at zero and infinity,
one can {\em regularize\/} the
integral by cutting out small neighborhoods of the two singularities
in such a way that the divergences cancel each other
and the remaining contributions are small.
Hence all
the exponents which yield opposite-signed divergence at both ends
can be steady solutions of Eq.~(\ref{KineticEquation}).
As we will see below this general statement helps 
to describe the experimental oceanographic data which are available to us.
The nature of such  steady solutions 
depends on the  particular truncation of the 
divergent integrals. 

So far, we have kept the formalism at the level of the self-similar
limit of the kinetic equation~(\ref{KineticEquation}).
However,
once one considers energy transfer mechanisms dominated by 
interactions with extreme modes of the system, one can no longer neglect
the deviations from self-similarity near the spectral boundaries:
the inertial frequency due to the rotation of the Earth at the IR end, and the buoyancy frequency and/or
dissipative cut-offs at the UV end.  

For example, we may consider a scenario in
which interactions with the smallest horizontal wavenumbers dominate 
the energy transfer within the inertial subrange, either because
the collision integral at infinity converges or
because the system is more heavily truncated at the large wavenumbers by
wave breaking or dissipation.
We will demonstrate
that the IR divergence of the collision integral
has a simple physical interpretation:
the evolution of each wave is dominated
by the interaction with its nearest neighboring vertical wavenumbers, mediated by
the smallest horizontal wavenumbers of the system. Such  a mechanism is denoted ``Induced Diffusion''
in the oceanographic literature.

To bring back the effects of the rotation of the Earth
in Eq.~(\ref{KineticEquation}), one introduces the Coriolis parameter $f$
there and in the linear dispersion relation.  
Since we are considering the evolution of
waves with frequency $\omega$ much larger than 
$f$, $f$ can be considered to be small.  However, since
the interaction with waves near $f$ dominates the energy transfer,
one needs to invert the
order in which the limits are taken, postponing making 
$f$ zero to the end.
This gives rise to an integral
that diverges like $f$ raised to a negative power smaller than $-1$,  
but multiplied by a prefactor that
vanishes if either $9-2a-3b=0$ or $b=0$.
These are the induced diffusion lines of steady state solutions, found
originally in \cite{mccomas-1981} as a diffusive approximation to the kinetic equation.  This family of stationary states does a reasonable job of explaining the gamut of observed variability.  The rigorous asymptotic analysis presented here clearly implies the Induced Diffusion family of stationary states makes sense only in the IR divergent sub-domain of $(a,b)$ space and we find that the data are located in this sub-domain.

 The present paper investigates in detail the parameter space $(a,b)$ of a general power-law spectrum (\ref{PowerLawSpectrum}), compares this parameter space to the ocean observations and give possible interpretation. Furthermore, the present study places the previously obtained ID curves~\citep{mccomas-1981} and Pelinovsky--Raevsky spectrum~\citep{pelinovsky1977wti} into a much wider context. Lastly, we present a general theoretical background that we are going to exploit for future studies. 

The paper is organized as follows.
Wave turbulence theory for the internal wave field and
the corresponding kinetic equation 
are briefly summarized in Section~\ref{sec:WTKE}  along with the
motivating observations.
We analyze the divergence of the kinetic equation in Section~\ref{sec:SI}.  
Section \ref{NovelSolution}
includes a special, convergent power-law solution that may account for the GM spectrum.
In Section~\ref{Balance} we introduce 
possible quasi-steady solutions  of the kinetic equation which are based on cancellations 
of two singularities.
Section~\ref{CorRegularization} shows that the IR divergence is dominated by induced diffusion,
and computes the family of power-law solutions
which arises from taking it into account.
We conclude in Section~\ref{sec:conclusion}.

\section{Wave turbulence theory for internal waves}
\label{sec:WTKE}

\subsection{Background and history}

The idea of using wave turbulence formalism to describe internal waves is
certainly not new; it dates to 
\cite{K66,K68}, with calculations of the kinetic
equations for oceanic spectra presented in \cite{olbers1976net, mccomas-1977-82, pomphrey1980dni}.
Various formulations have been developed for characterizing 
wave-wave interactions in  stratified wave turbulence
in the last four decades (see \cite{elementLPY} for 
a brief review, and \cite{caillol2000kea,H66,K66,K68,gm_lvov,lvov-2004-195,mccomas-1977-82,milder,MO75,O74,olbers1976net,pelinovsky1977wti,pomphrey1980dni,voronovich1979hfi} 
for details).  We briefly discuss the derivation of the kinetic equation and wave-wave interaction matrix elements
below 
in Eq.~(\ref{MatrixElement}). 
 
The starting point for the most extensive investigations has been a non-canonical Hamiltonian formulation in Lagrangian coordinates~\citep{mccomas-1981} that requires an unconstrained approximation in smallness of wave amplitude in addition to the assumption that nonlinear transfers take place on much longer time scales than the underlying linear dynamics.  Other work has as its basis a  formulation in Clebsch-like variables~\citep{pelinovsky1977wti} and a non-Hamiltonian formulation in Eulerian coordinates~\citep{caillol2000kea}.  Here we employ a canonical Hamiltonian representation in isopycnal coordinates~\citep{gm_lvov,lvov-2004-195} which, as a canonical representation, preserves the original symmetries and hence conservation properties of the original equations of motion.  

Energy transfers in the kinetic equation are characterized by three simple
mechanisms identified by
\cite{mccomas-1977-82} and reviewed by \cite{muller1986nia}.  These mechanisms 
represent extreme scale-separated limits.
One of these mechanisms represents the
interaction of two small vertical scale, high frequency waves with a
large vertical scale, near-inertial (frequency near $f$) wave and has
received the name Induced Diffusion (ID).  The ID mechanism exhibits a
family of stationary states, i.e.\ a family of solutions to
Eq.~(\ref{PowerLawSpectrum}).  A comprehensive inertial-range theory with
constant downscale transfer of energy can be obtained by patching
these mechanisms together in a solution that closely mimics the
empirical universal spectrum (GM) \citep{mccomas-1981}. A fundamental caveat from this work 
is that  the interaction time scales of high
frequency waves are sufficiently small at small spatial scales as to
violate the assumption of weak nonlinearity.

In parallel work, \cite{pelinovsky1977wti} derived a kinetic equation
for oceanic internal waves. They also have found the statistically steady state spectrum of internal waves, Eq.~(\ref{LT}),
which we propose to call Pelinovsky-Raevsky spectrum.
This spectrum was later found in \cite{caillol2000kea,gm_lvov,lvov-2004-195}.  
It follows from applying the Zakharov--Kuznetsov conformal transformation \cite{zak_book}, which effectively establishes a map between the very large and very small wavenumbers.  Making these two contributions cancel pointwise yields the solution~(\ref{LT}).

Both \cite{pelinovsky1977wti} and \cite{caillol2000kea} noted that
the solution~(\ref{LT}) comes through a cancellation between oppositely signed
divergent contributions in their respective collision integrals.  A
fundamental caveat is that one can not use conformal mapping for
divergent integrals.  Therefore, the existence of such a solution is
fortuitous.

Here we demonstrate that our canonical Hamiltonian structure admits  a similar characterization:  power-law solutions of the form
(\ref{PowerLawSpectrum}) return collision integrals that are, in general, divergent.  Regularization of the integral allows us to examine the
conditions under which it is possible to rigorously determine the power-law exponents $(a,b)$ in Eq.~(\ref{PowerLawSpectrum}) that lead to stationary states.  In doing so we obtain the ID family.

The situation is somewhat peculiar:  We have assumed
weak nonlinearity to derive the kinetic equation. The kinetic equation then predicts that nonlocal, strongly scale-separated interactions 
dominate the dynamics. These interactions have  less chance to 
be weakly nonlinear than regular, ``local'' interactions.  
Thus the derivation of the kinetic equation and its self-consistency
are at risk.  In our subsequent work \citep{elementLPY} we provide a possible
resolution of this puzzle. Yet,
as we will see below, despite this caveat, the weakly nonlinear
theory is consistent with much of the observational evidence.

\subsection{Experimental motivation}
\label{sec:PRL}
Power laws provide a simple and intuitive physical description of
complicated wave fields. Therefore we assumed that the spectral energy
density can be represented as Eq.~(\ref{PowerLawSpectrum}), and
undertook a systematic study of published observational programs. In
doing so we were fitting the experimental data available to us by
power-law spectra. We assume that the power laws offer a good
fit of the data in the high-frequency, high-wavenumer parts of the
spectrum. We do not  assume that spectra are given by Garrett and
Munk spectrum.

In most instances, vertical wavenumber and frequency power laws were estimated by superimposing best fit lines on top of one-dimensional spectra.  The quoted power laws are the asymptotic relations of these best fit lines.  Fits in the frequency domain included only periods smaller than 10 hours, thereby eliminating the inertial peak and semi-diurnal tides from consideration.  For one-dimensional spectra there is an implicit assumption that the
high-frequency-high-wavenumber spectra are separable.  Since both 1-d
spectra are red, frequency spectra are typically dominated by low
vertical wavenumber motions and vertical wavenumber spectra are
dominated by low frequencies.  Care should be taken to  distinguish
these results from the high wavenumber asymptotics of truly two-dimensional spectra.  
We have included two realizations of two-dimensional {\em displacement} spectra in isopycnal coordinates (Patchx$^{2}$ and Swapp).  The quoted power laws in these instances were estimated using a straight edge and $\chi$-by-eye procedure.  

Below we list extant data sets with concurrent vertical profile and
current meter observations and some major experiments utilizing moored
arrays, along with our best estimate
of their high-wave-number high frequency asymptotics:
\begin{itemize}
\item Site-D~\citep{Foff69, ST09}: 
Energy Spectra are $m^{-2.0}$ {\rm and} $\omega^{-2.0}$.
\item The Frontal Air-Sea Interaction Experiment
  (FASINEX)~\citep{weller1991for,eriksen1991ofv}: 
Energy Spectra are $m^{-2.3}$ {\rm and} $\omega^{-1.85}$. 
\item  The Internal Wave Experiment (IWEX)~\citep{M78}: 
Energy Spectrum  is - $k^{-2.4 \pm 0.4}\omega^{-1.75}$.
\item Salt Finger Tracer Release Experiment (SFTRE)~\citep{Schmitt05} / Polymode IIIc (PMIII)~\citep{Keffer83};  
Energy Spectrum are $m^{-2.4}$ {\rm and} $\omega^{-1.9}$. 
\item North Atlantic Tracer Release Experiment (NATRE)~\citep{P03} / Subduction~\citep{Weller04}: 
Energy spectra are $m^{-2.55}$ (observed, NATRE$^1$) or $m^{-2.75}$ (minus vortical contribution, NATRE$^{2}$) and $\omega^{-1.35}$. 
\item The Patches Experiment (PATCHEX$^1$)~\citep{Gregg93,chereskin}: 
Energy Spectra are $m^{-1.75}$ {\rm and} $\omega^{-1.65} ~-~\omega^{-2.0}$.
\item The Patches Experiment (PATCHEX$^2$)~\citep{sherman21evw}: 
Energy Spectrum is $m^{-1.75}\omega^{-1.65} ~-~m^{-1.75}\omega^{-2.0}$.
\item  The Surface Wave Process Program (SWAPP) experiment~\citep{anderson1992ssa}:  
Energy Spectrum is $m^{-1.9}\omega^{-2.0}$. 
\item Storm Transfer and Response Experiment (STREX)~\citep{DA84} /
  Ocean Storms Experiment (OS)~\citep{DA95}: 
Energy Spectra  are $m^{-2.3}$ {\rm and} $\omega^{-2.2}$.
\item Midocean Acoustic Transmission Experiment (MATE), \cite{L86}:  
Energy Spectra are { - $m^{-2.1}$ {\rm and} $\omega^{-1.7}$}. 
\item The Arctic Internal Wave Experiment
  (AIWEX)~\citep{Letal87,dasaro1991iwa}: 
Energy Spectra are $m^{-2.25}$ {\rm and} $\omega^{-1.2}$.
\end{itemize}
Two estimates of the Natre spectrum are provided:  Natre$^{1}$ represents the observed spectrum, Natre$^{2}$ represents the observed spectrum minus
the quasi-permanent finestructure spectrum identified in \cite{P03}.  The residual (Natre$^{2}$) represents our 'best' estimate of the internal wave
spectrum.  Two estimates of the Patchex spectrum are provided:  Patchex$^{1}$ combines free-fall vertical profiler data from \cite{Gregg93} and
long-term current meter data from \cite{chereskin}, Patchex$^2$ is an estimate from a two-dimensional displacement spectrum appearing in \cite{sherman21evw}.
Further details and a regional characterization of these data appear in \cite{iwobsPL}.

Finally, power laws of a two dimensional vertical wavenumber - frequency spectrum, $e(m,\omega) \propto \omega^{-c} m^{-d}$, correspond to the power laws of a three dimensional vertical wavenumber - horizontal wavenumber action spectrum $n(k,m) \propto k^{-a}m^{-b}$ with the mapping:
$$
a = c + 2   {\rm ~~and~~   }  b = d - c .  
$$

\begin{figure}
 \begin{center}
\includegraphics[scale=0.6]{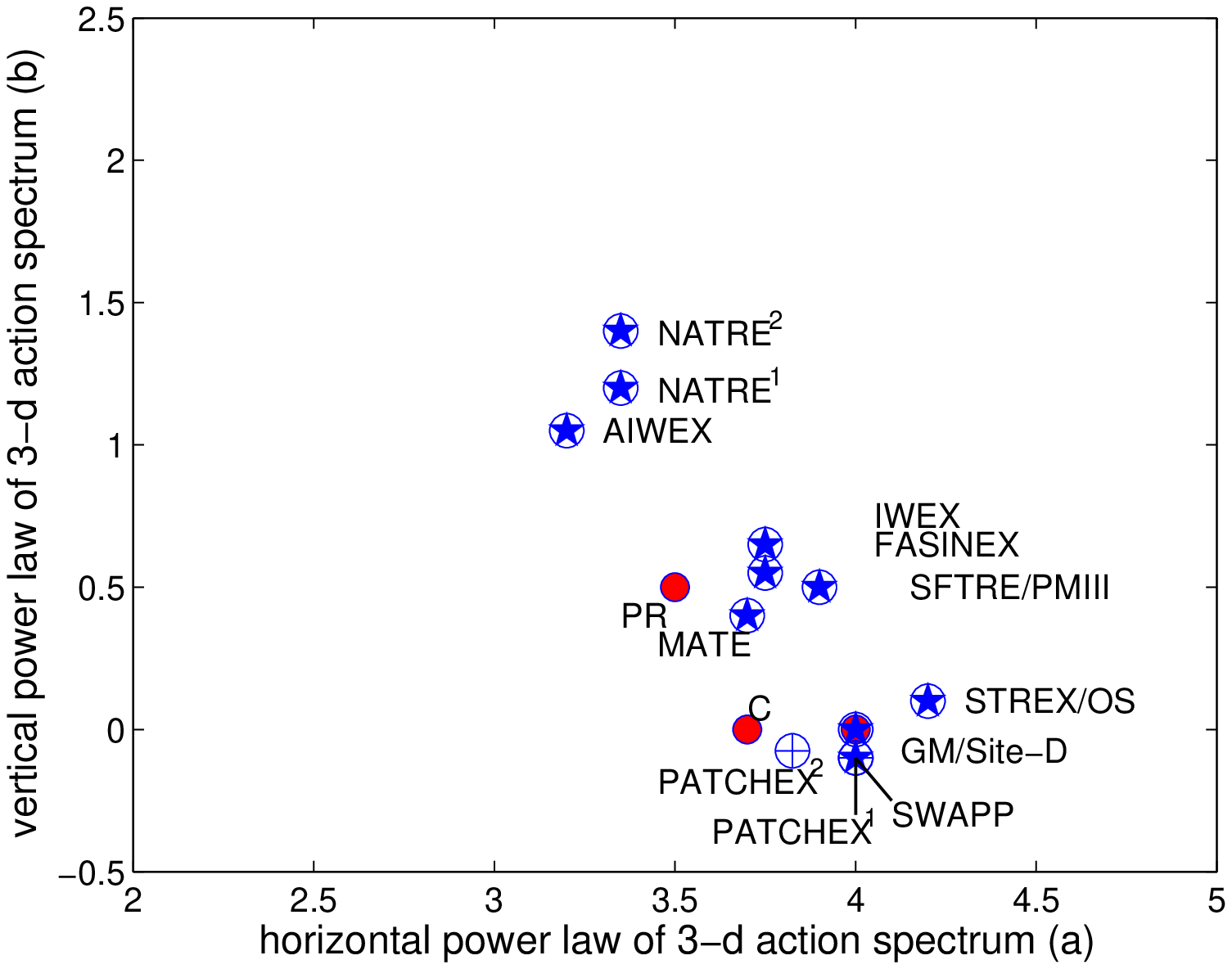}
 \end{center}
\caption{The observational points.  The filled circles represent the Pelinovsky--Raevsky (PR) spectrum[$(a,b) = (3.5,0.5)$], the convergent numerical solution determined in Section~\ref{NovelSolution} (C)[$(a,b) = (3.7,0.0)$] and the GM spectrum[$(a,b) = (4.0,0.0)$].  Circles with stars represent estimates based upon one-dimensional spectra from the western North Atlantic south of the Gulf Stream (IWEX, FASINEX and SFTRE/PMIII), the eastern North Pacific (STREX/OS and PATCHEX$^1$), the western North Atlantic north of the Gulf Stream (Site-D), the Arctic (AIWEX) and the eastern North Atlantic (NATRE$^1$ and NATRE$^2$).  There are two estimates obtained from two-dimensional data sets from the eastern North Pacific (SWAPP and PATCHEX$^2$) represented as circles with cross hairs.   NATRE$^1$ and NATRE$^{2}$ represent fits to the observed spectra and observed minus vortical mode spectra, respectively. Therefore NATRE$^2$ represents the ``best'' estimate of the NATRE internal wave spectrum. To conclude, he figure contains 12 observational points from 10 observational programs (as two programs produced two points each). Note that PATCHEX$^1$ is indistinguishable from SWAPP. Also note that one of the three filled circles (GM) coincides with the experimental point from Site-D.    }
\label{fig:onlydata}
\end{figure}

Figure (\ref{fig:onlydata}) suggests that the data points are not
randomly distributed, but have some pattern.
Explaining the location of the experimental points and making sense out of this pattern is the main physical motivation for this study.

\subsection{Hamiltonian Structure and Wave Turbulence Theory}
\label{HamiltonianSection}

This subsection briefly summarizes the derivation in 
\cite{gm_lvov,lvov-2004-195}; it is included 
here only for completeness and to allow references from
the core of the paper.

The equations of motion satisfied by an
incompressible stratified rotating flow in hydrostatic balance
under the Boussinesq approximation are \citep{Cushman}:
\begin{eqnarray}
\frac{\partial}{\partial t}\frac{\partial z}{\partial \rho} + \nabla \cdot \left(\frac{\partial z}{\partial \rho} \bm{u} \right) &=& 0 , \nonumber \\
\frac{\partial \bm{u}}{\partial t} +f \bm{u}^\perp+ \bm{u} \cdot
\nabla \bm{u} + \frac{\nabla M}{\rho_0} &=& 0 ,
\nonumber
\\
\frac{\partial M}{\partial \rho} - g z &=& 0 .
\label{PrimitiveEquations}
\end{eqnarray}
These equations result from mass conservation, horizontal momentum conservation and hydrostatic balance.
The equations are written in isopycnal coordinates with the density
$\rho$ replacing the height $z$ in its role as independent vertical
variable.  Here $\bm{u} = (u,v)$ is the horizontal component of the
velocity field, $\bm{u}^{\perp} = (-v, u)$, $\nabla = (\partial/\partial x, \partial/\partial y)$ is the gradient
operator along isopycnals,
$M$ is the Montgomery potential
 $$M=P+g\,\rho\,z \, , $$ 
$f$ is the Coriolis parameter, 
and $\rho_0$ is a reference density in its role as
inertia, considered
constant under the Boussinesq approximation.

The potential vorticity is given by
\begin{equation}
 q = \frac{f+\partial v/\partial x - \partial u/\partial y}{\Pi} ,
\label{PVorig}
\end{equation}
where
$\Pi = \rho / g \partial^2 M/\partial \rho^2 = \rho \partial z/\partial \rho $
is a normalized differential layer thickness. Since both the potential vorticity
and the fluid density
are conserved along particle trajectories,
an initial profile of the potential
vorticity that is a function of the density will be preserved by the flow.
Hence it is self-consistent to assume that
\begin{equation}
  q(\rho) = q_0(\rho) = \frac{f}{\Pi_0(\rho)} \, ,
 \label{PV}
\end{equation}
where $\Pi_0(\rho) = -g / N(\rho)^2$ is a reference stratification profile
with constant background buoyancy frequency, $N = (-g/(\rho \partial z/\partial\rho|_{\mathrm{bg}}))^{1/2}$, independent of $x$ and $y$. This assumption
is not unrealistic: it represents a pancake-like distribution of potential vorticity,
the result of its comparatively faster homogenization along than across isopycnal surfaces.

It is shown in \cite{gm_lvov,lvov-2004-195} that the
primitive equations of motion (\ref{PrimitiveEquations})
under the assumption ({\ref{PV}) 
can be written as a pair of
canonical Hamiltonian equations,
\begin{equation}
 \frac{\partial \Pi}{\partial t}  = - \frac{\delta {\cal H}}{\delta \phi} \, ,
  \qquad
\frac{\partial \phi}{\partial t}  = \frac{\delta {\cal H}}{\delta \Pi} \, ,
  \label{Canonical}
\end{equation}
where $\phi$ is the isopycnal velocity potential, and
the Hamiltonian is the sum of kinetic and potential energies,
%
\begin{eqnarray}
 {\cal H} = \!\! \int \!\! d \bm{x} d \rho
 \left(
 - \frac{1}{2} \left(
\Pi_0+\Pi(\bm{x}, \rho) \right) \,
  \left|\nabla \phi(\bm{x}, \rho) + \frac{f}{\Pi_0} \nabla^{{\perp}}\Delta^{-1} \Pi(\bm{x}, \rho)
  \right|^2
+
  \frac{g}{2} \left|\int^{\rho} d\rho^{\prime} \frac{\Pi(\bm{x}, \rho^{\prime})}{\rho^{\prime}} \right|^2
  \right) .
\nonumber\\
  \label{HTL2}
\end{eqnarray}
%
Here, $\nabla^{\perp}=(-\partial / \partial y, \partial / \partial x)$, $\Delta^{-1}$ is the inverse Laplacian and $\rho^{\prime}$ represents a variable of integration rather than perturbation.  

Switching to Fourier space, and
introducing a complex field variable $c_{\bm{p}}$ through the transformation
\begin{eqnarray}
 \phi_{\bm{p}} &=& \frac{i N \sqrt{\omega_{\bm{p}}}}{\sqrt{2 g} |\bm{k}|} \left(c_{\bm{p}}-
c^{\ast}_{-{\bm{p}}}\right)\, ,\nonumber\\
\Pi_{\bm{p}} &=& \Pi_0-\frac{N\, \Pi_0\,
|\bm{k}|}{\sqrt{2\, g \omega_{\bm{p}}}}\left(c_{\bm{p}}+c^{\ast}_{-{\bm{p}}}\right)\, ,
\label{transformationToSingleEquation2}
\end{eqnarray}
where the frequency $\omega$ satisfies the linear dispersion relation 
\footnote{This dispersion relation is written in the isopycnal framework. 
In the more familiar Eulerian framework, the dispersion relation
transforms into
$$ 
\omega_{\bm{p}}=\sqrt{f^2+\frac{N^2 k^2 }{m_{\ast}^2}},
$$
where $m_{\ast}$, the vertical wavenumber in $z$ coordinates, is given by
$m_{\ast} = -\frac{g}{\rho_0 N^2} m$ .}
\begin{eqnarray}
\omega_{\bm{p}}=\sqrt{ f^2 + \frac{g^2}{\rho_0^2 N^2} \frac{|\bm{k}|^2 }{m^2}},
\label{InternalWavesDispersion}
\end{eqnarray}
the equations of motion
(\ref{PrimitiveEquations}) adopt the canonical form
\begin{equation}
i\frac{\partial}{\partial t} c_{\bm{p}} = \frac{\delta {\cal H}}{\delta
c_{\bm{p}}^{\ast}} \, ,\label{fieldequation}
\end{equation}
with Hamiltonian:
\begin{eqnarray}
&& {\cal H} = \int d\bm{p} \, \omega_{\bm{p}} |c_{\bm{p}}|^2
\nonumber \\
 && 
\quad
+ \int d\bm{p}_{012}
\left(
 \delta_{\bm{p}+\bm{p}_1+\bm{p}_2} (U_{\bm{p},\bm{p}_1,\bm{p}_2} c_{\bm{p}}^{\ast} c_{\bm{p}_1}^{\ast} c_{\bm{p}_2}^{\ast} + \mathrm{c.c.})
+ \delta_{-\bm{p}+\bm{p}_1+\bm{p}_2} (V_{\bm{p}_1,\bm{p}_2}^{\bm{p}} c_{\bm{p}}^{\ast} c_{\bm{p}_1} c_{\bm{p}_2} + \mathrm{c.c.})
\right)\nonumber\\
.
\label{HAM}
\end{eqnarray}
This is the standard form of the Hamiltonian of a system dominated by 
three-wave interactions~\citep{zak_book}.
Calculations of interaction coefficients are tedious but straightforward task, completed in \cite{gm_lvov,lvov-2004-195}.
These coefficients are  given by 
\begin{subequations}
 \begin{eqnarray}
&&
V_{\bm{p}_1,\bm{p}_2}^{\bm{p}} = \frac{N}{4 \sqrt{2g}} 
  \frac{1}{k k_1 k_2}
  \left(
  I_{\bm{p},\bm{p}_1,\bm{p}_2} + J_{\bm{p}_1,\bm{p}_2}^{\bm{p}} + K_{\bm{p},\bm{p}_1,\bm{p}_2}
  \right) ,
  \label{eq:V}
 \\
&&
U_{\bm{p},\bm{p}_1,\bm{p}_2} = \frac{N}{4 \sqrt{2g}} \frac{1}{3}
  \frac{1}{k k_1 k_2}
  \left(
  I_{\bm{p},\bm{p}_1,\bm{p}_2} + J_{\bm{p}_1,\bm{p}_2}^{-{\bm{p}}} + K_{\bm{p},\bm{p}_1,\bm{p}_2}
  \right),
\label{eq:U}
  \\
&&
I_{\bm{p},\bm{p}_1,\bm{p}_2} =
 - \sqrt{\frac{\omega_1 \omega_2}{\omega}}
  k^2 \bm{k}_1 \cdot \bm{k}_2
  - \left( (0,1,2) \rightarrow (1,2,0) \right)
  - \left( (0,1,2) \rightarrow (2,0,1) \right)
,
\label{eq:I}
\\
&&
J_{\bm{p}_1,\bm{p}_2}^{\bm{p}} = \frac{f^2}{\sqrt{\omega \omega_1 \omega_2}}
  \big(k^2 \bm{k}_1 \cdot \bm{k}_2
  - \left( (0,1,2) \rightarrow (1,2,0) \right)
  - \left( (0,1,2) \rightarrow (2,0,1) \right)
\big)
,
\label{eq:J}
\\
&&
K_{\bm{p},\bm{p}_1,\bm{p}_2} = -i f
\bigg(
  \sqrt{\frac{\omega}{\omega_1 \omega_2}}
  (k_1^2 - k_2^2)
  \bm{k}_1 \cdot \bm{k}_2^{\perp}
\nonumber\\
&&
\qquad\qquad\qquad\qquad
  + \big( (0,1,2) \rightarrow (1,2,0) \big)
  + \big( (0,1,2) \rightarrow (2,0,1) \big)
\bigg)
,
\label{eq:K}%
 \end{eqnarray}%
\label{MatrixElement}%
\end{subequations}
where $((0,1,2) \to (1,2,0))$ and $((0,1,2) \to (2,0,1))$ denote exchanges of suffixes, and for two dimensional vector $\bm{k} = (k_x,k_y)$, 
$\bm{k}^{\perp} =  (-k_y,k_x)$ \footnote{We note that these are correct expressions, which coincide
with those given in \cite{lvov-2004-195}, apart from a couple of $1/2$
factors.}.
We stress that the field equation
(\ref{fieldequation}) with the three-wave Hamiltonian
(\ref{InternalWavesDispersion}, \ref{HAM}, \ref{MatrixElement}) is {\em
equivalent\/} to the primitive equations of motion for internal waves
(\ref{PrimitiveEquations}).  The approach using a Lagrangian coordinate system is based on small-amplitude expansion to arrive to this type of equation. 

In wave turbulence theory, one proposes a perturbation
expansion in the amplitude of the nonlinearity, yielding 
linear waves at the leading order. Wave amplitudes are
modulated by the nonlinear interactions,  
and the modulation is statistically described by a kinetic equation~\citep{zak_book} for
the wave action $n_{\bm{p}}$ defined by
\begin{eqnarray}
n_{\bm{p}} \delta(\bm{p} - \bm{p}^{\prime}) = \langle c_{\bm{p}}^{\ast} c_{\bm{p}^{\prime}}\rangle.
\label{WaveAction}
\end{eqnarray}
Here $ \langle \dots \rangle$ denotes an ensemble averaging, i.e. averaging over many realizations of the random wave field. 
The derivation of this kinetic equation is well studied and understood~\citep{zak_book,lvov2004nsl}.
For the three-wave Hamiltonian (\ref{HAM}), the kinetic equation
is the one in
Eq.~(\ref{KineticEquation}),
describing general internal waves
interacting in both rotating and non-rotating environments. 

The delta functions in the kinetic equation ensures that spectral transfer happens on the 
{\em resonant manifold\/}, defined as 
\begin{eqnarray}
\mathrm{(a)}~~
\left\{
\begin{array}{l}
 \bm{p} = \bm{p}_1 + \bm{p}_2 \\
 \omega = \omega_1 + \omega_2
\end{array}
\right. ,
\qquad
\mathrm{(b)}~~
\left\{
\begin{array}{l}
 \bm{p}_1 = \bm{p}_2 + \bm{p} \\
 \omega_1 = \omega_2 + \omega
\end{array}
\right. ,
\qquad
\mathrm{(c)}~~
\left\{
\begin{array}{l}
 \bm{p}_2 = \bm{p} + \bm{p}_1 \\
 \omega_2 = \omega + \omega_1
\end{array}
\right. .
\label{RESONANCES}
\end{eqnarray}

Now let us assume that the wave action $n$ is independent of the direction of the 
horizontal wavenumber, and is symmetric with respect to $m \to-m $ change
$$n_{\bm{p}} = n(|\bm{k}|, |m|) .$$ Note  that value of the interaction matrix element is independent of horizontal azimuth as it depends only on the magnitude of interacting wavenumbers.  
Therefore one can integrate
the kinetic equation (\ref{KineticEquation}) 
over horizontal azimuth~\citep{zak_book}, yielding
\begin{eqnarray}
&& \frac{\partial n_{\bm{p}}}{\partial t}
= \frac{2}{k}\int
\left(R^{0}_{12} - R^{1}_{20} - R^{2}_{01} \right) \,
d k_1 d k_2 d m_1 d m_2 \, , \nonumber \\
&&
R^{0}_{12}=
   f^{\bm{p}}_{\bm{p}_1\bm{p}_2} \, |V^{\bm{p}}_{\bm{p}_1\bm{p}_2}|^2 \, \delta_{m-m_1-m_2} 
\delta_{\omega_{\bm{p}}-\omega_{\bm{p}_1}-\omega_{\bm{p}_2}}
k k_1 k_2
/ S^0_{1,2}
\, .
 \label{KEinternalAveragedAngles}
\end{eqnarray}
Here $ S^0_{1,2} $ appears as the result of integration of the horizontal-momentum conservative delta function over all 
possible orientations and is equal to the area of the triangle 
with sides  with the length
of the horizontal wavenumbers
$k = |\bm{k}|$, $k_1 = |\bm{k}_1|$ and $k_2  = |\bm{k}_2|$.
This is the form of the kinetic equation which will be used to find scale-invariant solutions in the next section.

\section{Scale-invariant kinetic equation}
\label{sec:SI}

\subsection{Reduction of Kinetic Equation to the Resonant Manifold}
\label{Reduction}

In the high-frequency limit $\omega\gg f $, 
one could conceivably neglect the
effects of the rotation of the Earth.
The dispersion relation (\ref{InternalWavesDispersion}) then becomes~\citep{gm_lvov}
\begin{equation}
\omega_{\bm{p}} \equiv \omega_{\bm{k},m} \simeq
 \frac{g}{\rho_0 N} \frac{|\bm{k}|}{|m|} \, ,
\label{HighFrequencyDispersion}
\end{equation}
and, to the leading order, the matrix element
(\ref{MatrixElement}) retains only its first term, $I_{\bm{p},\bm{p}_1,\bm{p}_2}$.

The azimuthally-integrated kinetic equation~(\ref{KEinternalAveragedAngles}) includes integration over $k_1$ and $k_2$ 
since the integrations over $m_1$ and $m_2$ can be done by using delta functions. To use delta functions, we need to perform what is called reduction to the resonant manifold. 
Consider, for example, resonances of type (\ref{RESONANCES}a). Given
$k$, $k_1$, $k_2$
and $m$, one can find $m_1$ and $m_2$ satisfying
the resonant condition by solving simultaneous equations
\begin{eqnarray}
m = m_1 + m_2 ,
\qquad
\frac{k}{|m|}=\frac{k_1}{|m_1|} +\frac{k_2}{|m-m_1|} .
\label{ResonantConditionM2}
\end{eqnarray}
The solutions of this quadratic equation are given by
\begin{subequations}
 \begin{eqnarray}
&&
\left\{
\begin{array}{l}
m_1 = \displaystyle\frac{m}{2 k} \left(k + k_1 + k_2 + \sqrt{(k + k_1 + k_2)^2 - 4 k k_1}\right)
\\
m_2 = m - m_1
\end{array}
\right.
,
\label{eq:sol1}
\\
  \mathrm{and} \nonumber \\
&&
\left\{
\begin{array}{l}
m_1 = \displaystyle\frac{m}{2k} \left(k - k_1 - k_2 - \sqrt{(k - k_1 - k_2)^2 + 4 k k_1}\right)
\\
m_2 = m - m_1
\end{array}
\right.
.
\label{eq:sol2}
\end{eqnarray}
\label{eq:sol12}
\end{subequations}
Note that Eq.~(\ref{eq:sol1}) translates into Eq.~(\ref{eq:sol2}) if the indices $1$ and $2$ are exchanged. 
Indeed, exchanging indices $1$ and $2$ in Eq.~(\ref{eq:sol1}) we obtain
 \begin{eqnarray}
m_1^{\prime} = m - m_2^{\prime} = m - \displaystyle\frac{m}{2 k} \left(k + k_1 + k_2 + \sqrt{(k + k_1 + k_2)^2 - 4 k k_2}\right),\nonumber
\end{eqnarray}
which simplifies then to  $m_1$ of Eq.~(\ref{eq:sol2}).

Similarly, resonances of type (\ref{RESONANCES}b) yield
\begin{subequations}
 \begin{eqnarray}
&&
\left\{
\begin{array}{l}
m_2 = - \displaystyle\frac{m}{2 k} \left(k - k_1 - k_2 + \sqrt{(k - k_1 - k_2)^2 + 4 k k_2}\right)
\\
m_1 = m + m_2
\end{array}
\right.
,
\label{eq:sol3}
\\
\mathrm{and} \nonumber\\
&&
\left\{
\begin{array}{l}
m_2 = - \displaystyle\frac{m}{2k} \left(k + k_1 - k_2 + \sqrt{(k + k_1 - k_2)^2 + 4 k k_2}\right)
\\
m_1 = m + m_2
\end{array}
\right.
.
\label{eq:sol4}
 \end{eqnarray}
\label{eq:sol134}
\end{subequations}
and resonances of type (\ref{RESONANCES}c) yield
\begin{subequations}
 \begin{eqnarray}
&&
\left\{
\begin{array}{l}
m_1 = - \displaystyle\frac{m}{2k} \left(k - k_1 - k_2 + \sqrt{(k - k_1 - k_2)^2 + 4 k k_1}\right)
\\
m_2 = m + m_1
\end{array}
\right.
,
\label{eq:sol5}
\\
\mathrm{and} \nonumber \\
&&
\left\{
\begin{array}{l}
m_1 = - \displaystyle\frac{m}{2k} \left(k - k_1 + k_2 + \sqrt{(k - k_1 + k_2)^2 + 4 k k_1}\right)
\\
m_2 = m + m_1
\end{array}
\right.
.
\label{eq:sol6}
 \end{eqnarray}
\label{eq:sol56}
\end{subequations}

After this reduction, a double integral
over $k_1$ and $k_2$ is left. The domain of integration is further restricted by the
triangle inequalities
\begin{equation}
 k<k_1+k_2, \quad k_1<k+k_2, \quad \mathrm{and} \quad k_2<k+k_1.
\label{KinematicBoxK}
\end{equation}
These conditions ensure that one can construct a triangle out of the wavenumbers with lengths $k$, $k_1$ and $k_2$ and determine the domain
in the $(k_1,k_2)$ plane called the {\em kinematic box\/} in the oceanographic literature.

Numerical evaluation of the collision integral is a complicated yet
straightforward task.  Interpretation of the results, though, is more
difficult, mostly due to the complexity of the interaction matrix element
and the nontrivial nature of the resonant set.  Starting with
\cite{mccomas-1977-82}, therefore, predictions were made based on a
further simplification. This simplification is based on the assertion that
it is interactions between wavenumbers with extreme scale separation
that contribute mostly to the 
nonlinear dynamics.
Three main  classes of such resonant triads appear, characterized by extreme
scale separation. These three main classes are 
\begin{itemize}
\item the vertical backscattering of a high-frequency wave
      by a low-frequency wave of twice the vertical wavenumber
      into a second high-frequency wave of oppositely signed vertical wavenumber.
      This type of scattering, as in 
      Eqs.~(\ref{eq:ES1}, \ref{eq:ES2}, \ref{eq:ES1inf}, \ref{eq:ES2inf}) below, 
      is called elastic scattering (ES).
\item The scattering of a high-frequency wave by a low-frequency, small-wavenumber wave into a second, nearly identical, high-frequency large-wavenumber wave.
      This type of scattering, as in 
      Eqs.~(\ref{eq:ID1}, \ref{eq:ID2}, \ref{eq:ID1inf}, \ref{eq:ID2inf}) below,
      is called induced diffusion (ID).
\item The decay of a small-wavenumber wave into two large vertical-wavenumber waves of approximately one-half the frequency.
      This type of scattering, as in 
      Eqs.~(\ref{eq:PSI1}, \ref{eq:PSI2}, \ref{eq:PSI1inf}, \ref{eq:PSI2inf}) below,
      is called parametric subharmonic instability (PSI).
\end{itemize}

To see how this classification appears analytically, we
perform the limit of $k_1 \to 0$ and the limit $k_1\to\infty$ in
Eqs~(\ref{eq:sol12}--\ref{eq:sol56}). We will refer to the $k_1$ {\em or\/} $k_2 \to 0$ limits as 
IR (Infrared) limits, while the $k_1$ {\em and\/} $k_2 \to \infty$ limit
will be referred as an UV (Ultra Violet) limit.  Since the integrals in the kinetic
equation for power-law solutions will be dominated by the scale-separated
interaction, this will help us analyze possible solutions to the kinetic equation.

The results of the $k_1\to  0 $ limit of 
Eqs.~(\ref{eq:sol12}--\ref{eq:sol56}) are given by
\begin{subequations}
 \begin{eqnarray}
&
\left\{
\begin{array}{l}
m_1 \to 2m, m_2 \to -m
\\
\omega_1 \ll \omega, \omega_2 \sim \omega
\end{array}
\right.
,
\label{eq:ES1}
\\
&
\left\{
\begin{array}{l}
-m_1 \ll m, m_2 \sim m
\\
\omega_1 \ll \omega, \omega_2 \sim \omega
\end{array}
\right.
,
\label{eq:ID1}
 \end{eqnarray}
\end{subequations}
\begin{subequations}
 \begin{eqnarray}
&
\left\{
\begin{array}{l}
m_1 \ll m, m_2 \sim -m
\\
\omega_1 \sim 2\omega, \omega_2 \sim \omega
\end{array}
\right.
,
\label{eq:PSI1}
\\
&
\left\{
\begin{array}{l}
-m_1 \ll m, m_2 \sim -m
\\
\omega_1 \sim 2\omega, \omega_2 \sim \omega
\end{array}
\right.
,
\label{eq:PSI2}
\end{eqnarray}
\end{subequations}
\begin{subequations}
 \begin{eqnarray}
&
\left\{
\begin{array}{l}
-m_1 \ll m, m_2 \sim m
\\
\omega_1 \ll \omega, \omega_2 \sim \omega
\end{array}
\right.
,
\label{eq:ID2}
\\
&
\left\{
\begin{array}{l}
m_1 \to -2m, m_2 \to -m
\\
\omega_1 \ll \omega, \omega_2 \sim \omega
\end{array}
.
\right.
\label{eq:ES2}
\end{eqnarray}
\end{subequations}
We now see that 
the interactions (\ref{eq:ES1}, \ref{eq:ES2}) correspond
to the elastic scattering (ES) mechanism,  
the interactions (\ref{eq:ID1}, \ref{eq:ID2}),
correspond to the induced diffusion (ID).  The interactions (\ref{eq:PSI1}, \ref{eq:PSI2}),
correspond to the  parametric subharmonic instability (PSI).

Similarly, taking the $k_1$ and $k_2 \to \infty$ limits, of 
Eqs.~(\ref{eq:sol12}--\ref{eq:sol56}) we obtain
\begin{subequations}
 \begin{eqnarray}
&
\left\{
\begin{array}{l}
m_1 \gg m, -m_2 \gg m
\\
\omega_1, \omega_2 \sim \omega/2
\end{array}
\right.
,
\label{eq:PSI1inf}
\\
&
\left\{
\begin{array}{l}
-m_1 \gg m, m_2 \gg m
\\
\omega_1, \omega_2 \sim \omega/2
\end{array}
\right.
,
\label{eq:PSI2inf}
 \end{eqnarray}
\end{subequations}
\begin{subequations}
 \begin{eqnarray}
&
\left\{
\begin{array}{l}
m_1 \sim m/2, m_2 \sim -m/2
\\
\omega_1,\omega_2 \gg \omega
\end{array}
\right.
,
\label{eq:ES1inf}
\\
&
\left\{
\begin{array}{l}
-m_1 \gg m, -m_2 \gg m
\\
\omega_1,\omega_2 \gg \omega
\end{array}
\right.
,
\label{eq:ID1inf}
\end{eqnarray}
\end{subequations}
\begin{subequations}
 \begin{eqnarray}
&
\left\{
\begin{array}{l}
m_1 \sim -m/2, m_2 \sim m/2
\\
\omega_1, \omega_2 \gg \omega
\end{array}
\right.
,
\label{eq:ES2inf}
\\
&
\left\{
\begin{array}{l}
-m_1 \gg m, -m_2 \gg m
\\
\omega_1, \omega_2 \gg \omega
\end{array}
.
\right.
\label{eq:ID2inf}
\end{eqnarray}
\end{subequations}
We now can identify the interactions~(\ref{eq:PSI1inf}, \ref{eq:PSI2inf}) to be PSI, 
the interactions~(\ref{eq:ES1inf}, \ref{eq:ES2inf}) to be ES,
and finally the interactions~(\ref{eq:ID1inf}, \ref{eq:ID2inf}) as being ID.

This classification provides an easy and intuitive tool for describing extremely scale-separated interactions. We will see below that one of these interactions, namely ID, explains reasonably well the experimental data that is available to us.

\subsection{Convergences and divergences  of the kinetic equation}
\label{ssec:tw}

Neglecting the effects of the rotation of the Earth
yields a scale-invariant system
with dispersion relation given by Eq.~(\ref{HighFrequencyDispersion})
and matrix element given only by the $I_{\bm{p},\bm{p}_1,\bm{p}_2}$ in Eq.~(\ref{MatrixElement}).
This is the kinetic equation of \cite{gm_lvov,lvov-2004-92,lvov-2004-195}, describing
internal waves in hydrostatic balance in a non-rotating environment.

Proposing a self-similar separable spectrum of the form
(\ref{PowerLawSpectrum}), one can show from 
the azimuthally-integrated kinetic equation 
(\ref{KEinternalAveragedAngles}) that~\citep{zak_book}
\begin{equation}
\frac{\partial n(\alpha \bm{k}, \beta m)}{\partial t} = \alpha^{4 - 2 a} \beta^{1 - 2b} \frac{\partial n(\bm{k} ,m)}{\partial t}
\label{Rescaling}
\end{equation}
for constants $\alpha$ and $\beta$.  In order to 
find a steady scale-invariant solution for all values of $k$
and $m$, it is therefore sufficient to find exponents that give zero collision integral for one wavenumber.
One can fix $k$ and $m$, adopting for instance $k=m=1$, and seek zeros of the collision integral (represented as ${\cal C}$ below) as a 
function of $a$ and $b$:
\begin{equation}
 \frac{\partial n(k=1,m=1)}{\partial t}\equiv  {\cal C}(a,b) .
\label{ScaleInvariantKineticEquation}
\end{equation}

\begin{figure}
 \begin{center}
  \includegraphics[]{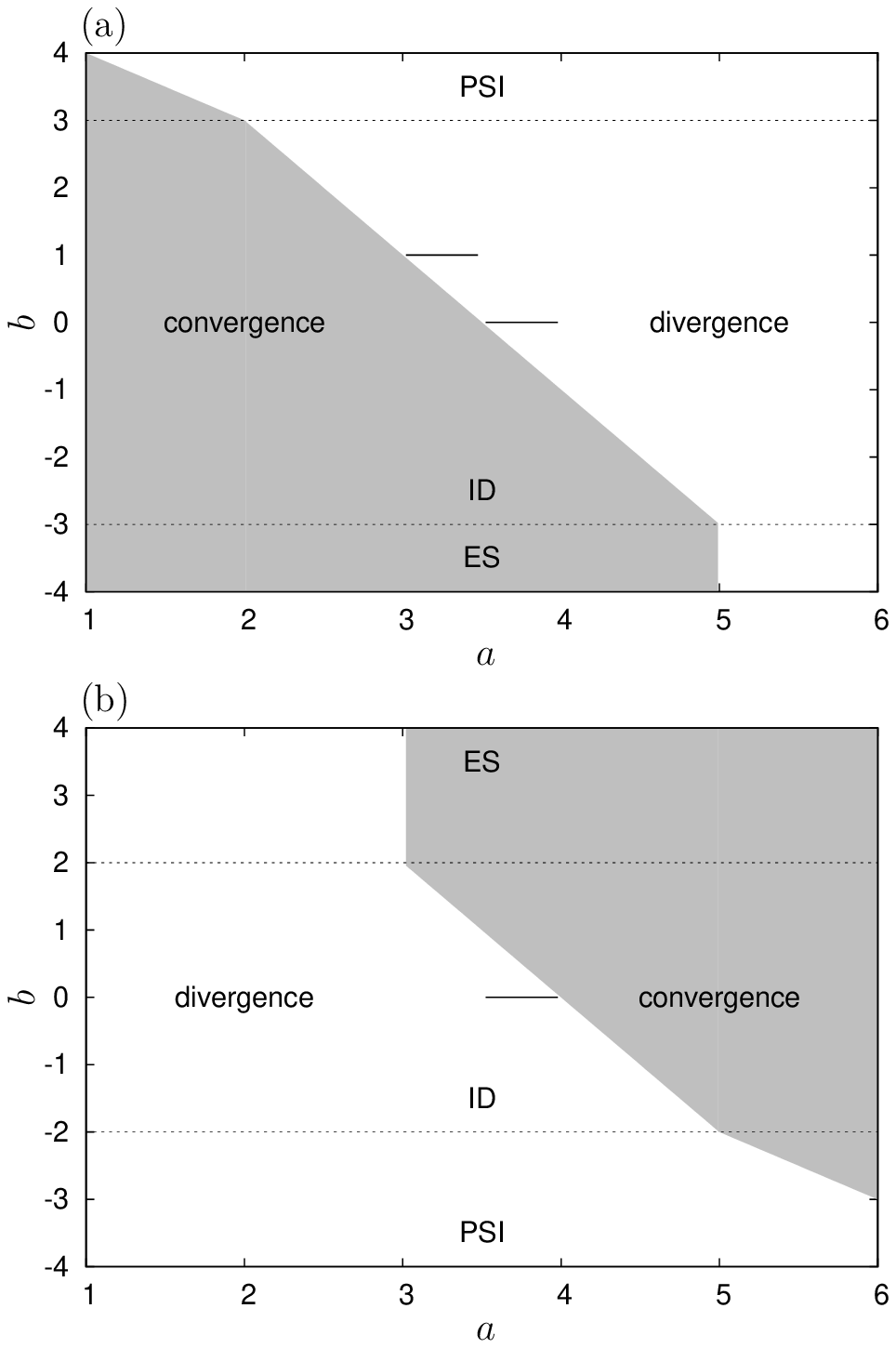}
 \end{center}
 \caption{Divergence/convergence due to IR wavenumbers (a)
and due to UV wavenumbers (b).
The integral converge  for the exponents in the shaded regions
or on the segments $(b=0, 7/2 < a < 4$) and $(b=1, 3 < a < 7/2)$.  Dashed lines distinguish the 
domains where the indicated named triads dominate the singularity.}
\label{DivergenceConvergence}
\end{figure}

Before embarking on numerical or analytical integration of the kinetic equation 
(\ref{KEinternalAveragedAngles}) with scale-invariant solutions~(\ref{PowerLawSpectrum}), it is 
necessary to check whether or not the collision integral converges.
Appendix A
outlines these calculations. The condition for the scale-invariant collision integral~(\ref{ScaleInvariantKineticEquation}) 
to converge at the IR end, $k_1 \;\mathrm{or}\; k_2 \to 0$ is given by
\begin{subequations}
 \begin{eqnarray}
 &
a+b/2-7/2<0 \;\; \mathrm{and} \;\; -3 < b < 3, \label{IR1}
\\
&
a-4<0 \;\; \mathrm{and} \;\; b=0,              \label{IR2}
\\
&
a-7/2<0 \;\; \mathrm{and} \;\; b=1 ,           \label{IR3}
\\
&
a+b-5<0 \;\; \mathrm{and} \;\; b>3 ,            \label{IR4}
\\
\mathrm{or} \nonumber\\
&
a-5<0 \;\; \mathrm{and} \;\; b<-3 .                \label{IR5}
\end{eqnarray}
\label{InfraRed}
\end{subequations}

Similarly, UV convergence as $k_1 \;\mathrm{and}\; k_2 \to \infty$ implies that
\begin{subequations}
 \begin{eqnarray}
 &
a+b/2-4>0 \;\; \mathrm{and} \;\; -2<b<2, \label{UV1}
\\
&
a-7/2>0 \;\; \mathrm{and} \;\; b=0 ,      \label{UV2}
\\
&
a-3>0 \;\; \mathrm{and} \;\; b>2 ,         \label{UV3}
\\
\mathrm{or} \nonumber\\
&
a+b-3>0 \;\; \mathrm{and} \;\; b<-2 .      \label{UV4}
\end{eqnarray}
\label{UltraViolet}
\end{subequations}
The domains of divergence and convergence
are shown in Fig.~\ref{DivergenceConvergence}.

Figure~\ref{DivergenceConvergence} 
also displays the classes of 
triads dominating the interactions.
Knowing the classes of interactions that lead to the divergences of the kinetic
equation allows us to find possible physical scenarios of the 
convergent solutions or to find a possible physical regularization of the 
divergences.

Note that in addition to the two-dimensional
domain of IR convergence [the regions~(\ref{IR1}, \ref{IR4}, \ref{IR5})] there are
two additional IR convergent line segments given by Eqs.~(\ref{IR2}) and
(\ref{IR3}). These two special line segments appear because 
of the $b(b-1)$ prefactor
to the divergent contributions to the collision integral~(\ref{ZeroContribution}).  Similarly, for the UV limit, in addition to 
the two-dimensional region of convergence~(\ref{UV1}, \ref{UV3}, \ref{UV4})
there is an additional special line segment of $b=0$~(\ref{UV2}).

We see that these domains of convergence overlap only on the segment
\begin{equation}
 7/2 < a < 4 \;\; \mathrm{and} \;\; b = 0 \, . \label{ConvergentSolution}
\end{equation}
Note that  $b=0$ corresponds to wave action independent of vertical wavenumbers, 
$\partial n / \partial m = 0$.
Existence of the $b=0$ line will allow us to find novel convergent solution 
in Section~\ref{NovelSolution}. 
We also note that the IR segment on $b=0$ coincides with one of the ID 
solution determined in Section~\ref{CorRegularization}.  
The other segment on $b=1$ do not coincides with the ID solution in Section~\ref{CorRegularization} since the scale-invariant system has higher symmetry
than the system with Coriolis effect.

\section{A novel convergent solution \label{NovelSolution}}
To find out whether there is a steady solution
of the kinetic equation along the convergent segment~(\ref{ConvergentSolution}), we
substitute the power-law ansatz~(\ref{PowerLawSpectrum}) with $b=0$
into the azimuthally-integrated kinetic equation~(\ref{KEinternalAveragedAngles}).
We then compute numerically the
collision integral as a function of $a$ for $b=0$. To this end, 
we fix $k=m=1$ and perform a numerical integration over
the kinematic box (\ref{KinematicBoxK}), reducing the integral to the resonant
manifold as described in Subsection~\ref{sec:SI}\ref{Reduction}.

The result of this numerical integration is shown in 
Fig.~\ref{ThreeAndSeven}. The figure clearly  shows the  
existence of a steady solution
of the kinetic equation~(\ref{KEinternalAveragedAngles})
near $a \cong 3.7$ and $b=0$.

\begin{figure}
\begin{center}
 \includegraphics[]{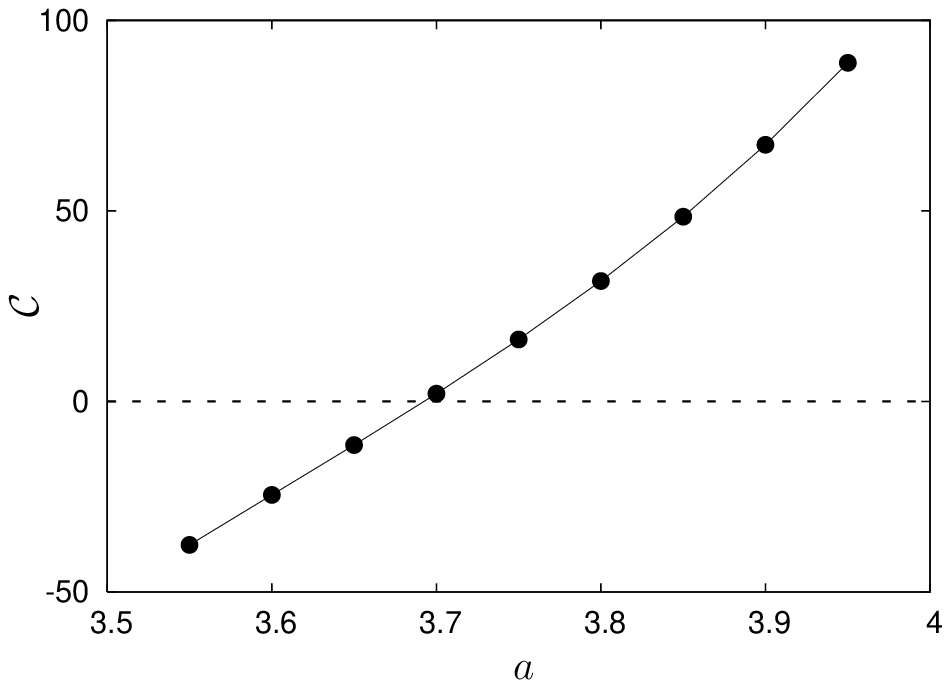}
\end{center}
\caption{Value of the collision integral as a function of $a$
on the convergent segment, $b=0$.}
\label{ThreeAndSeven}
\end{figure}
This is, therefore, {\em the only convergent steady solution\/} to the scale-invariant kinetic
equation for the internal wave field. It is highly suggestive that
it should exist so close to the GM spectrum, $a=4$ and $b=0$ for large wavenumbers, the most
agreed-upon fit to the spectra observed throughout the ocean. It remains to be 
seen whether and how this solution is modified by inclusion of background rotation.

\section{Balance between divergences\label{Balance}}

The fact that the collision integral ${\cal C}$
diverges for almost
all values of $a$ and $b$
can be viewed both as a challenge and as a blessing. On the one hand,
it makes the prediction of steady spectral slopes more difficult,
since it now depends on the details of the truncation of the domain of
the integration.
Fortunately,
it provides a powerful tool for quantifying
the effects of fundamental players in ocean dynamics, most of which
live on the fringes of the inertial subrange of the internal wave field:
the Coriolis effect, as well as tides and storms at the IR end
of the spectrum, and wave breaking and dissipation at the UV
end. The sensitive response of the inertial subrange to the detailed
modeling of these scale-separated mechanisms permits, in principle, building
simple models in which these are the only players, bypassing the need to
consider the long range of wave scales in between.

\begin{figure}
 \begin{center}
  \includegraphics[]{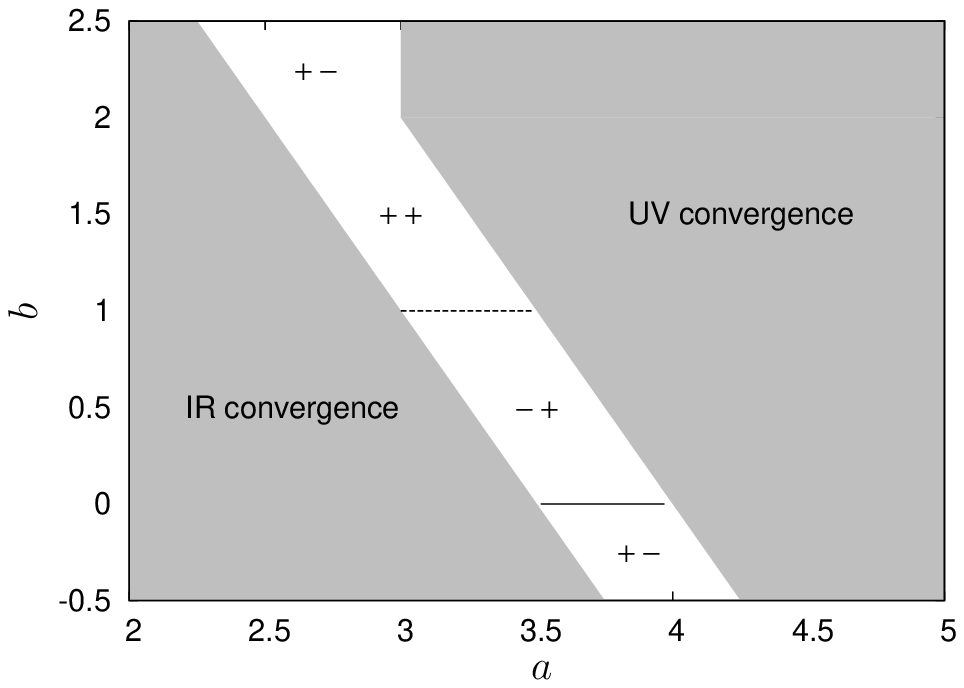}
 \end{center}
 \caption{Signs of the divergences
 where the both IR and UV contributions diverge.
 The left symbols show the signs due to IR wavenumbers and
 the right symbols show the signs due to UV wavenumbers.}
\label{SignContributions}
\end{figure}

At the IR end,
the resonant interactions are dominated by the ID singularity for $-3<b<3$, 
Fig.~\ref{DivergenceConvergence}.  
The sign of the divergences is given by $-b(b-1)$
[Eq.~(\ref{ZeroContribution})].
Similarly,
at the UV end resonant interactions are dominated by 
the ID singularity for $-2<b<2$ (Fig.~\ref{DivergenceConvergence}).
The sign of the divergence is given by $b$ (Eq.~(\ref{InfinityContribution})).
At the UV end for $b>2$,
where ES determines the divergences,
the sign of the singularity is given by $-b$, i.e. the sign is negative
(Eq.~(\ref{InfinityContributionES})).
Figure~\ref{SignContributions} shows 
the signs of the divergences
where both the IR and UV contributions diverge:
the left sign corresponds to the IR contribution,
and the right sign to the UV contribution.

Hence,
in the regions,
\begin{subequations}
 \begin{eqnarray}
&
 7 < 2a+b < 8, \;\;\mathrm{and}\;\; -2 < b < 1
,
\\
\mathrm{or} \nonumber\\
&
7 < 2a+b, a<3, \;\;\mathrm{and}\;\; b>2
,
\end{eqnarray}
\end{subequations}
the divergences of the collision integral at the
IR and UV ends have opposite signs. 
Then formal solutions can be constructed by 
having these two divergences cancel each other out. 

This observation justifies the existence of the PR solution (\ref{LT}). Indeed, the  PR spectrum
has divergent power-law exponents at the both ends.
One can prove that
the PR spectrum is an exact steady solution of the kinetic equation~(\ref{KineticEquation})
by applying the Zakharov--Kuznetsov conformal
mapping for systems with cylindrical symmetry~\citep{zak1968,zakharov1967iwn,kuznetsov1972tis}.
This Zakharov--Kuznetsov conformal mapping
effectively establishes a map between the
neighborhoods of zero and infinity.  Making these two contributions
cancel pointwise yields the solution~(\ref{LT}).  For this transformation to be mathematically 
applicable the integrals have to converge.  This transformation leads
only to a formal solution for divergent integrals.
Some other transformation, as we explain below, may lead to completely different solutions.

The PR spectrum was first found by \cite{pelinovsky1977wti}.
However,
they realized that it was only a formal solution.  
The solution was found again in \cite{caillol2000kea} through a
renormalization argument, and in \cite{gm_lvov,lvov-2004-195}
within an isopycnal formulation of the wave field.

The idea of a formal solution, such as PR, can be generalized 
quite widely: in fact,
{\em any\/} point in the regions with opposite-signed divergences
can be made into a
steady solution
under a suitable conformal mapping that makes the divergences at zero and infinity cancel each
other, as does the Zakharov--Kuznetsov transformation. 
Such {\em generalized\/} Zakharov--Kuznetsov transformation 
is an extension of the idea of principal value for a divergent integral,
whereby two divergent contributions are made to cancel each other through a
specific relation between their respective contributions.  

Indeed, in the ocean internal waves can not have zero horizontal
wavenumber. Rather, there exists a smallest horizontal wavenumber,
which corresponds to largest horizontal scales of internal waves.  The
largest wavenumber that was observed is on the order of thousands of
kilometers. On larger scales $\beta$ effects become important, and
they prevent wavenumbers to achieve even smaller scales. Other effects
possibly affecting small wavenumbers include ocean storms,
interactions with large scale vortices and shear, as well as 
 ocean boundaries.  
Similarly there is no infinitely large wavenumber for internal waves,
rather there is possibly large horizontal wavenumber that is affected
by wave breaking, interaction with the turbulence, other processes.

The idea of a generalized Zakharov--Kuznetsov transformation leading to an infinite number of steady states provides a possible explanation for the variability of the power-law exponents of the quasi-steady spectra: inertial subrange spectral variability is to be expected
when it is driven by the nonlocal interactions. The natural local
variability of players outside the inertial range  translates into a certain degree of non-universality.
Such players include storms and tides, as well as possible geometrical constrains, interactions with 
large scale shear and vortices. 
Investigation of the nature of possible balances between IR and UV
divergences is outside of scope of the present manuscript. 

\section{Regularization by the Coriolis effect}
\label{CorRegularization}

Physically, the ocean does not perform  
generalized Zakharov--Kuznetsov transformation.  
However, in the ocean there are finite boundaries in the frequency 
domain. In particular, 
the inertial frequency, $f$, provides a truncation
for the IR part of the spectrum, while the UV truncation is provided by the buoyancy frequency, $N$.  
These two frequencies vary from place to place, giving grounds for
spectral variability.
Consequently, the
integrals are not truly divergent, rather they have a large
numerical value dictated by the location of the IR cutoff. 

Note that the $f \to 0$ calculations presented in this section describes 
an intermediate frequencies $\omega$ such that $\omega\gg f>0$. Consequently, 
these intermediate frequencies feel inertial frequency as being small. 
These calculations will not be applicable for equatorial regions, where $f$ is 
identically zero.

Observe that all the experimental points are located in regions of the $(a,b)$ domain $a > (-b + 7)/2$ 
for which the integral {\it diverges} in the IR region.  
Also note that five out of twelve experimental points are located in the region where collision integrals are UV {\it convergent}, i.e. in  $a > -b/2 + 4$.   
The UV region is therefore assumed to be 
either sub-dominant or convergent in this section, where 
we study the regularization resulting from a finite value of $f$. 

We note that because we consider scale-invariant case only in the
present manuscript, the IR cutoff can equivalently be considered as
$k$ approaching smallest possible value, or equivalently, when
$\omega$ approaches its smallest value. 

Since the IR cutoff is given by $f$, a frequency,
it is easier to analyze the resulting integral in $(\omega_1,m_1)$
rather than in the more traditional $(k_1,m_1)$ domain utilized in \citep{muller1986nia} and the  previous sections.
We emphasize that the present paper concerns itself with {\it scale-invariant}
wave action spectrum (\ref{PowerLawSpectrum}). The scale-invariant waveaction 
can be translated from $(k_1,m_1)$  to  $(\omega_1,m_1)$ domain with out 
loss of generality.  This statement would not be true for the realistic
Garrett and Munk spectrum, or any other oceanic spectrum due to the 
non-scale invariant form of the linear dispersion relation 
(\ref{InternalWavesDispersion}). Such spectra will be analyzed in subsequent
publication \citep{elementLPY}.

Therefore to proceed, we assume a power-law spectrum, similar to Eq.~(\ref{PowerLawSpectrum}), but in the
$(\omega, m)$ space:
\begin{eqnarray}
n_{\omega,m} \propto \omega^{\widetilde{a}} m^{\widetilde{b}}.
\label{PowerLawSpectrumFrequencyVertical}
\end{eqnarray}
We need to transform the wave action as a function of $k$ and $m$
to a function of $\omega$ and $m$.
This is done in
Appendix B.
The relation between $a$, $b$ and
$\widetilde{a}$, $\widetilde{b}$ reads:
$$\widetilde{a}=-a, \quad \widetilde{b}=-a-b.$$

  Thus we need to 
express both the kinetic equation and the kinematic box in 
terms of frequency and vertical wavenumber.
For this, we use the dispersion relation (\ref{InternalWavesDispersion}) 
to express $k$ in terms of $\omega$ in the description of the kinematic box
(\ref{KinematicBoxK}): 
\begin{subequations}
 \begin{eqnarray}
&& \omega_1<E_3(m_1), \; \omega_1>E_4(m_1), \; \omega_1>E_1(m_1)
\quad \mathrm{if\ } m_1<0, \; \omega_1>\omega,
 \label{Region1}\\
&& \omega_1>E_3(m_1), \; \omega_1<E_4(m_1), \; \omega_1<E_1(m_1)
 \quad \mathrm{if\ } m_1<0, \; \omega_1<\omega,
 \label{Region2}\\
&& \omega_1<E_2(m_1), \; \omega_1>E_2(m_1), \; \omega_1>E_4(m_1)
 \quad \mathrm{if\ } m_1>0, \; \omega_1<\omega,
 \label{Region3}\\
&& \omega_1>E_3(m_1), \; \omega_1<E_1(m_1), \; \omega_1<E_2(m_1)
\quad \mathrm{if\ } m_1>0, \; \omega_1>\omega,
 \label{Region4}%
\end{eqnarray}%
 \label{Region}%
\end{subequations}
where we have introduced the four curves in the $(\omega_1,m_1)$ domain that
parametrize the kinematic box:
\begin{eqnarray}
E_1(\omega_1) &=& m \frac{
 -\sqrt{-f^2+\omega^2} +    \sqrt{-f^2+ (\omega-\omega_1)^2}}
{ \sqrt{-f^2+\omega_1^2} +  \sqrt{-f^2+(\omega-\omega_1)^2}} , \nonumber\\
E_2(\omega_1) &=& m \frac{
  \sqrt{-f^2+\omega^2} +    \sqrt{-f^2+ (\omega-\omega_1)^2}}
{ \sqrt{-f^2+\omega_1^2} +  \sqrt{-f^2+(\omega-\omega_1)^2}}, \nonumber\\
E_3(\omega_1) &=& m \frac{
 -\sqrt{-f^2+\omega^2} +    \sqrt{-f^2+ (\omega-\omega_1)^2}}
{-\sqrt{-f^2+\omega_1^2} +  \sqrt{-f^2+(\omega-\omega_1)^2}} , \nonumber\\
E_4(\omega_1) &=& m \frac{
  \sqrt{-f^2+\omega^2} +    \sqrt{-f^2+ (\omega-\omega_1)^2}}
{ - \sqrt{-f^2+\omega_1^2} +  \sqrt{-f^2+(\omega-\omega_1)^2}} . \nonumber
 \label{KinematicBoxOM}
\end{eqnarray}

\begin{figure}
\begin{center}
\includegraphics[]{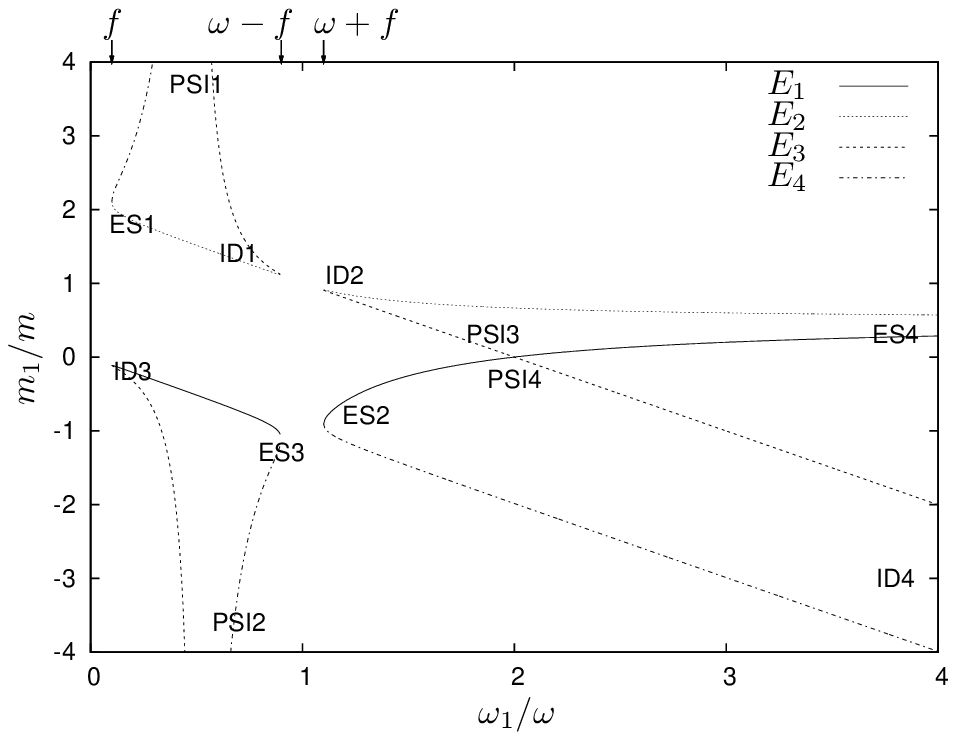}
\end{center}
\caption{The kinematic box in the $(\omega_1,m_1)$ domain. 
Two disconnected regions where $\omega_1 < \omega$ depict regions with
``sum'' interactions, namely
$\omega=\omega_1+\omega_2,$
$m=m_1+m_2,$
type of the resonances.
The connected regions where $\omega_1 > \omega$ depict ``difference'' resonances,
$\omega_2=\omega_1-\omega$,
$m_2=m_1-m$.
The parameters are chosen so that  $f/\omega=0.1$. Frequency $f$ is marked on the top of the graph, and frequency $N$ is outside of the region of frequencies shown on  the graph.}
\label{pic:OmegaMKinematicBox}
\end{figure}

The {\em kinematic box} in the $(\omega,m)$ domain is shown in Fig.~\ref{pic:OmegaMKinematicBox}. 
Note that the region (\ref{Region2}) and (\ref{Region3}) can be transferred to 
each other by interchanging indices 1 and 2, consequently two disconnected
$\omega_1<\omega$ regions look like mirrored and shifted copies of each other.  
To help in the transition from the traditional kinematic box~(\ref{KinematicBoxK})
 to the kinematic box in $(\omega,m)$ domain, 
the following limits were identified:
\begin{itemize}
\item  ID1  is the ID  limit of Eq.~(\ref{eq:ID1}) with indices 1 and 2 being flipped
\item  ID2  is the ID  limit of Eq.~(\ref{eq:ID2}) with indices 1 and 2 being flipped
\item  ID3  is the ID  limit of Eq.~(\ref{eq:ID1}) 
\item  ID4  is the ID  limit of Eqs.~(\ref{eq:ID1inf}, \ref{eq:ID2inf})
\item  PSI1 is the PSI limit of Eq.~(\ref{eq:PSI1inf})
\item  PSI2 is the PSI limit of Eq.~(\ref{eq:PSI2inf})
\item  PSI3 is the PSI limit of Eq.~(\ref{eq:PSI1})
\item  PSI4 is the PSI limit of Eq.~(\ref{eq:PSI2})
\item  ES1  is the ES  limit of Eq.~(\ref{eq:ES1})
\item  ES2  is the ES  limit of Eq.~(\ref{eq:ES2}) with indices 1 and 2 being flipped
\item  ES3  is the ES  limit of Eq.~(\ref{eq:ES2}) with indices 1 and 2 being flipped
\item  ES4  is the ES  limit of Eq.~(\ref{eq:ES1inf})
\end{itemize}
An advantage of the $(\omega,m)$ presentation for the kinematic box is that 
it allows a transparent  reduction to the resonant manifold. A disadvantage is the 
curvilinear boundaries of the box, requiring more sophisticated analytical treatment. 

Equation~(\ref{KEinternalAveragedAngles}) transforms into
\begin{eqnarray}
\frac{\partial}{\partial t} n(k(\omega,m),m) &=&\frac{1}{k}\int d\omega_1 d m_1 J
              \frac{|V^0_{12}|^2}{S^0_{1,2}}
\left(n_1 n_2 - n(n_1+n_2)\right) 
\vline_{\small { \omega_2=\omega-\omega_1, m_2 = m - m_1}}
\nonumber\\
&-&\frac{2}{k}\int d\omega_1 d m_1 J\frac{|V^1_{02}|^2}{S^1_{2,0}}
\left(n n_2 - n_1 ( n + n_2)\right)
\vline_{\omega_2=\omega_1-\omega, m_2 = m_1 - m}
\label{KE3}
\end{eqnarray}
%
We have used the dispersion relation 
$
k_i=m_i\sqrt{\omega_i^2 - f^2},
$ 
and defined $J$ as the Jacobian of the transformation from $(k_1,k_2)$ into 
$(\omega_1,\omega_2)$, times the $k k_1 k_2 $ factor,
$$J= k k_1 k_2 \frac{d k_1}{d \omega_1} \frac{d k_2}{d \omega_2}.$$

In Fig.~\ref{pic:OmegaMKinematicBox}, there are three ID regions (or corners)  with significant
contribution to the collision integral in the IR limit:
\begin{enumerate}
\item ID1 region. In this region, $m_1$ is slightly bigger than $m$, $\omega_1$ slightly smaller than
$\omega$, and $\omega_2$ and $m_2$ are both very small. 
This region can be obtained from
the region (\ref{Region2}) above by interchanging indices $1$ and $2$. 
In this region, the waveaction $n_2$ is much smaller than waveaction $n$ and $n_1$:
$$n_2\gg n, \, n_1.$$  
\item ID2 region. In this region, $\omega_1$ is slightly bigger than $\omega+f$ where
$\omega_2=\omega_1-\omega$ is small, and $m_2=m_1-m$ is negative and small. This is 
the region (\ref{Region4}). Also
$$n_2\gg n, \, n_1.$$  
\item ID3 region. Small $\omega_1$, small negative $m_1$. This corresponds
to the  region (\ref{Region2}), where waveaction obeys
$$n_1\gg n, \, n_2.$$
Note that this region can be obtained from the region ID1 by flipping indices 
$1$ and $2$. Consequently, only one of the ID1 and  ID3 should be taken into account,
with a factor of two multiplying the respective contribution.  This also can be
seen from Fig.~\ref{pic:OmegaMKinematicBox} as ID1 and ID3 regions are
shifted mirror images of each other. 
\end{enumerate} 

Making these simplifications, and taking into account the areas of
integration in the kinematic box, we obtain
%
\begin{eqnarray}
\frac{\partial}{\partial t} n(k(\omega,m),m) &=& \frac{2}{k}
  \int _f^{f+\omega_{\mathrm{s}}} d \omega_1 \int_{E_3(\omega_1)}^{E_1(\omega_1)}
d m_1
J\frac{|V^0_{1,2}|^2}{S^0_{1,2}} n_1 (n_2 - n)
\nonumber\\
&&- {\frac{2}{k}} \int _{\omega-f-\omega_{\mathrm{s}}}^{\omega-f} d \omega_1
\int_{E_3(\omega_1)}^{E_1(\omega_1)} d m_1
J\frac{|V^2_{1,0}|^2}{S^2_{1,0}} n_2 (n - n_1)
,
\label{KE4}
\end{eqnarray}
%
where the small parameter $\omega_{\mathrm{s}}$ is introduced to restrict the integration
to a neighborhood of the ID corners.  The arbitrariness of the small parameter will not affect 
the end result below.

To quantify the contribution of near-inertial waves to a $(\omega,m)$ mode, we write
$$\epsilon \sim f \ll \omega=1. $$
Subsequently, near the region ID3 of the kinematic box, we write
$$\omega_1 = f + \epsilon,$$
while near the ID2 corners of the kinematic box we write
$$\omega_1 = \omega+f+\epsilon.$$
We then expand the resulting analytical  expression (\ref{KE4})  in powers of
$\epsilon$ and $f$ without making any assumptions on relative sizes.
These calculations, including the integration over vertical wavenumbers  $m_1$,
are presented in
Appendix C.
The resulting expression for the kinetic equation is   given by
\begin{eqnarray}
&& \!\!\!\!\!
\frac{\partial}{\partial t} n(k(\omega,m),m) = \frac{\pi}{4k}
\left( \widetilde{a} - \widetilde{b} \right) 
 \left( \widetilde{a} - 3 \left( 3 + \widetilde{b} \right)  \right)
\nonumber\\
&& \!
\times
m^{5 + 2 \widetilde{b}} \omega^{-3 + \widetilde{a} - \widetilde{b}}
  \int_0^{\mu} d \epsilon 
{\left( \epsilon + f \right) }^{4 + \widetilde{a} + \widetilde{b}} 
(\epsilon^2+ 2 \epsilon f + 17 f^2) ~.
\label{KE6}
\end{eqnarray}
The integral over $\epsilon$ diverges at $\epsilon=0$, 
if $f=0$ and if $6+\widetilde{a} +\widetilde{b} > -1$.\footnote{Naturally this condition coincides with (\ref{IR1}).}

However, if we postpone taking $f=0$ limit, we see that 
the integral is zero to leading order if 
\begin{eqnarray}
 \widetilde{a} - 3\,\left( 3 + \widetilde{b} \right) =0
\quad \mathrm{or} \quad
 \widetilde{a} - \widetilde{b} = 0
\end{eqnarray}
or, in terms of
$a$ and $b$, 
\begin{eqnarray}
9-2a-3b=0
\quad \mathrm{or} \quad
b=0.
\label{NewIDcurve}
\end{eqnarray}

This is the family of power-law steady-state solutions to the kinetic
equations dominated by infra-red ID interactions. These steady states
are identical to the ID stationary states identified by \cite{mccomas-1977-82},
 who derived a diffusive approximation to
their collision integral in the infra-red ID limit.
Note that \cite{mccomas-1981} interpreted $b=0$ as a no action flux in
vertical wavenumber domain, while $9-2a-3b=0$ is a constant action
flux solution. 

We note that one can use eikonal approach to describe these
types of interactions \citep{Eikonal1,Eikonal2,muller1986nia,henyey1986energy}. Advantages of Eikonal
approach versus our scale-invariant analysis is that it allows to
consider not only scale-invariant interactions in separable spectrum.
The possible disadvantage of the eikonal approach is that construction of 
a transport theory is far less rigorous.  For a more  detailed discussion on the differences
between resonant interaction approximation and eikonal approach we refer
the reader to \cite{polzin-2004}.

What is presented in this section is a rigorous 
asymptotic derivation of Eq.~(\ref{NewIDcurve}).
These ID solutions helps us to
interpret observational data of Fig.~\ref{fig:onlydata} that is currently available to us.
The value added associated with a rigorous asymptotic derivation is
the demonstration that the ID stationary states are meaningful only in
the IR divergent part of the $(a,b)$ domain.

\section{Conclusions}
\label{sec:conclusion}

The results in this paper provide an interpretation of the
variability in the observed spectral power laws. Combining Figs.~\ref{fig:onlydata}, \ref{DivergenceConvergence} and
\ref{SignContributions} with Eqs.~(\ref{GM}), (\ref{LT}) and
(\ref{NewIDcurve}), produces the results shown in Fig \ref{fig:everything}.  

A non-rotating scale invariant analysis analysis provides two
subdomains in which the kinetic equation converges in either the UV or
the IR limit (light grey shading), two subdomains in which the kinetic
equation diverges in both UV and IR limits but with oppositely signed
values (dark grey shading) and a single domain with similarly signed
UV and IR divergences (black shading).  In this nonrotating analysis,
a stationary state is possible only for oppositely signed divergences,
i.e. within the dark grey shaded regions.  Six of the observational
points lie in IR and UV divergent subdomains, a seventh (Site-D) is on
the boundary with the IR divergent - UV convergent subdomain.  Two of
the observational points lie in the domain of IR and UV divergence
having similar signs, which does {\em not} represent a possible
solution in the non-rotating analysis.  These two points lie close to
the boundaries of the 'forbidden' (black shaded) region, and
subtracting the vortical contribution from one (Natre$^{1}$) returns a
'best' estimate of the internal wave spectrum (Natre$^{2}$) that lies
out side the 'forbidden' region.

All the data lie in an IR divergent regime and hence a regularization
of the kinetic equation is performed by including a finite lower
frequency of $f$.  This produces a family of stationary states, the
induced diffusion stationary states.  These stationary states collapse
much of the observed variability.  The exception is the Natre
spectrum.

\begin{figure}
 \begin{center}
\includegraphics[scale=0.6]{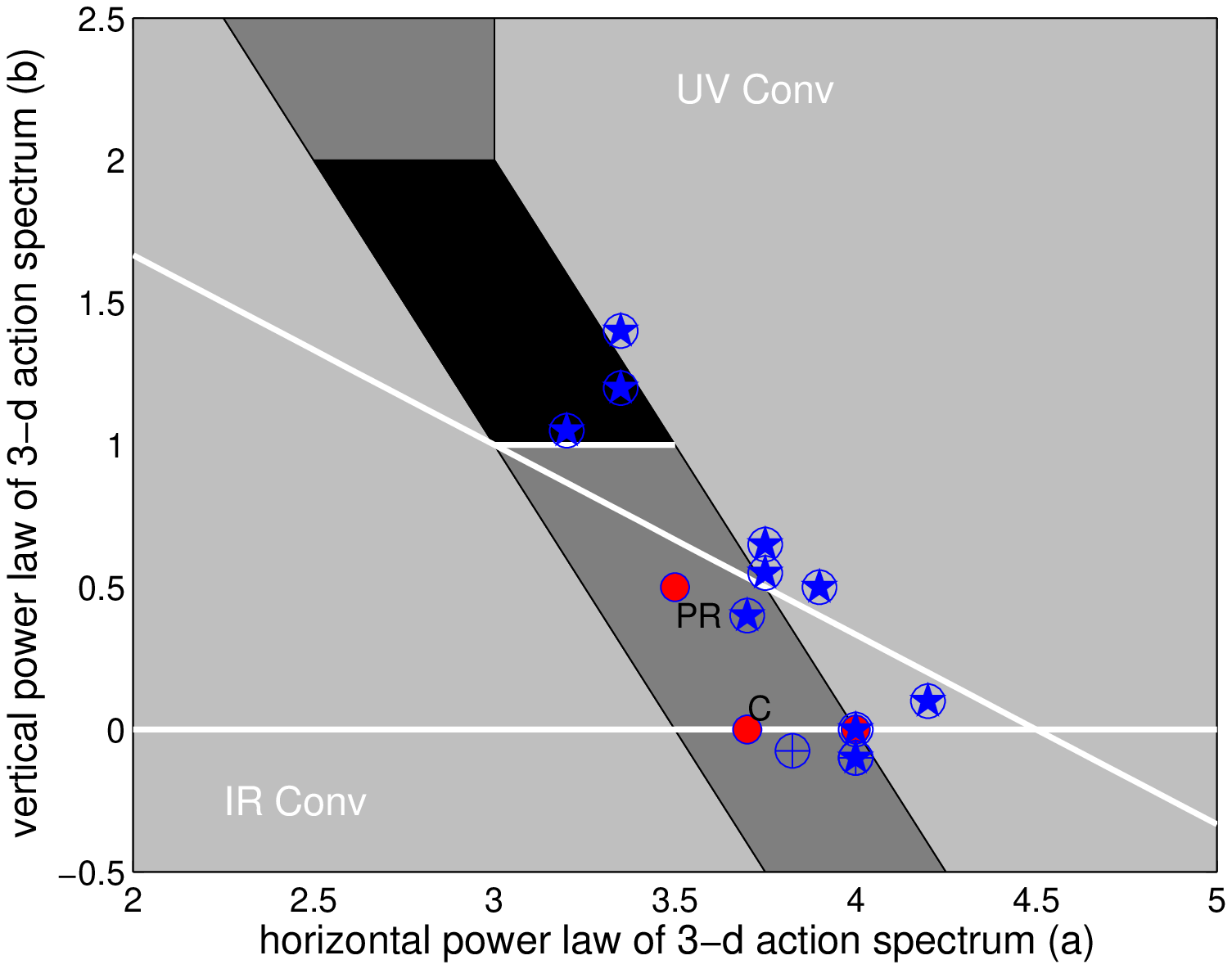}
 \end{center}
\caption{The observational points and the theories.  The filled circles represent the Pelinovsky--Raevsky (PR) spectrum, the convergent numerical solution determined in Section 4c and the GM spectrum.  Circles with stars represent power-law estimates based upon one-dimensional spectra.  Circles with cross hairs represent estimates based upon two-dimensional data sets.  See Fig.~\ref{fig:onlydata} for the identification of the field programs.  Light grey shading represents regions of the power-law domain for which the collision integral converges in either the IR or UV limit.  The dark grey shading represents the region of the power-law domain for which the IR and UV limits diverge and  have opposite signs.  The region of black shading represents the sub-domain for which both the IR and  UV divergences have the same sign, i.e. when large contributions from interactions with very small and very large wavenumbers have the same sign.  Overlain as solid white lines are the induced diffusion stationary states.  
}
\label{fig:everything}
\end{figure}

Summarizing the paper,
we have analyzed the scale-invariant kinetic equation for internal gravity waves,
and shown that its collision integral diverges
for almost all spectral exponents.
Figure~\ref{fig:everything} shows that
the integral nearly always diverges, either
at zero or at infinity or at both ends.
This means that, in  the wave turbulence kinetic equation framework, 
the energy transfer is dominated by the scale-separated interactions 
with  large and/or small scales.

The only exception where the integral converges
is a segment of a line, $7/2 < a < 4$, with $b=0$.
On this convergent segment, we found a special solution, $(a,b) = (3.7, 0)$. 
This new solution is not 
far from the large-wavenumber asymptotic form of the Garrett--Munk spectrum, $(a,b)=(4,0)$.

We have argued that there exist two sub-domains of power-law exponents
which can yield quasi-steady solutions of the kinetic equation.  
For these ranges of exponents,
the contribution of the scale-separated interactions
due to the IR and UV wavenumbers 
can be made to approximately balance each other.
The Pelinovsky--Raevsky spectrum is a special case of this scenario.

The scenario, in which the energy spectrum in the inertial subrange is determined 
by the nonlocal interactions, provides an explanation
for the variability of the power-law exponents of the observed spectra:
they are a reflection of the variability of dominant players outside
of the inertial range, such as
the Coriolis effect, tides and storms.

This possibility was further investigated by introducing rotation and then pursuing a rigorous asymptotic expansion of the kinetic equation.  In doing so we obtain the Induced Diffusion stationary states that appear as white lines in Fig.~\ref{fig:everything} which had previously been determined through a diffusive approximation.  Much of the observed oceanic variability lies about these stationary states in the IR divergent sub-domain.  

A more detailed review of available observational data used for this study appears
in~\cite{iwobsPL}. Numerical evaluation of the complete
non-scale-invariant kinetic equation of the Garrett--Munk spectrum is
prestented in \cite{elementLPY}, in which we also consider waves that
are slightly off resonant interactions.
The theory, experimental data, and results of numerical simulations in
\cite{numericsLY} all hint at the
importance of the IR contribution to the collision integral.  The nonlocal
interactions with large scales will therefore play a dominant role in
forming the internal-wave spectrum.  To the degree that the large
scales are location dependent and not universal, the high-frequency,
high vertical-wavenumber internal-wave spectrum ought to be affected
by this variability. 

Consequently, the internal-wave spectrum should be 
strongly dependent on the regional characteristics of the ocean, such as the
local value of the Coriolis parameter and specific features of the 
spectrum, specifically for near-inertial frequencies.

\begin{acknowledgment}
This research is supported by NSF CMG grants
0417724, 0417732 and 0417466. YL is also supported by NSF DMS grant 0807871 and ONR  Award N00014-09-1-0515.
We are grateful to YITP in Kyoto University for allowing us to use their facility.
\end{acknowledgment}

\begin{appendix}[A]\label{CalculationsOfDivergences}
\section*{Asymptotics of collision integral in infra-red and ultra-violet limits}
 
Let us integrate Eq.~(\ref{KEinternalAveragedAngles}) over $m_1$ and $m_2$.
\begin{eqnarray}
 \frac{\partial n_{\bm{p}}}{\partial t} &=&
\frac{1}{k}
  \int \left(T^0_{1,2} - T^1_{2,0} - T^2_{0,1}\right) dk_1 dk_2
,
\nonumber\\
 T^0_{1,2} &=& k k_1 k_2 |V^{\bm{p}}_{\bm{p}_1\bm{p}_2}|^2 f^{\bm{p}}_{\bm{p}_1\bm{p}_2}
  / (|g^{0 \prime}_{1,2}| S^0_{1,2})
,
\nonumber \\
 g^{0 \prime}_{1,2}(k_1, k_2) &=& \left.\frac{d g^0_{1,2}(m_1)}{d m_1} \right|_{m_1 = m_1^{\ast}(k_1, k_2)},
\qquad
 g^0_{1,2}(m_1) = \frac{k}{m} - \frac{k_1}{|m_1|} - \frac{k_2}{|m - m_1|} \, ,
\label{eq:keap}
\end{eqnarray}
where $g^{0 \prime}_{1,2}$ appears owing to $\delta_{\omega_{\bm{p}}
-\omega_{{\bm{p}_1}}-\omega_{{\bm{p}_2}}}$ and $m_1^{\ast}(k_1, k_2)$
is given by the resonant conditions~(\ref{eq:sol12}--\ref{eq:sol56}).

\subsection{Infra-red asymptotics}
We consider the asymptotics of the integral in Eq.~(\ref{eq:keap}) as $k_1 \to 0$.
We employ the independent variables $x$ and $y$,
where $k_1 = k x$, $k_2 = k (1 + y)$, $x,y = O(\epsilon)$,
$x>0$ and $-x<y<x$.
In this limit of $\epsilon \to 0$, $n_1 \gg n, n_2$.
In this limit,
Eqs.~(\ref{eq:sol1}) and (\ref{eq:sol6}),
Eqs.~(\ref{eq:sol2}) and (\ref{eq:sol5}),
and
Eqs.~(\ref{eq:sol3}) and (\ref{eq:sol4})
correspond to
ES, Eqs.~(\ref{eq:ES1}, \ref{eq:ES2}) 
ID, Eqs.~(\ref{eq:ID1}, \ref{eq:ID2}),
PSI, Eqs.~(\ref{eq:ES1}, \ref{eq:ES2}),
respectively.
Without loss of generality $m$ is set to be positive.

\begin{table*}
  \caption{Asymptotics as $k_1 \to 0$.
 ES (\ref{eq:sol1}, \ref{eq:sol6}) gives $\epsilon^{-a+5}$ owing to the symmetry of $y$.
 ID (\ref{eq:sol2}, \ref{eq:sol5}) gives $\epsilon^{-a-(b-7)/2}$ ($\epsilon^{-a+4}$) owing to the second cancellation.
PSI (\ref{eq:sol3}, \ref{eq:sol4}) gives $\epsilon^{-a-b+5}$.
The asymptotics for $b=0$ appear in parentheses.
}
 \label{table:zero}
 \begin{center}
\begin{tabular}{ccccccccc}
 & $m_1$ & $m_2$ & $\omega_1$ & $\omega_2$ & $V^{\bm{p}_i}_{\bm{p}_j\bm{p}_k}$ & $f^{\bm{p}_i}_{\bm{p}_j\bm{p}_k}$ & $g^{i \prime}_{j, k}$ & $T^i_{j,k}$
\\
\hline
\hline
(\ref{eq:sol1}) & $\epsilon^{0}$ & $\epsilon^{0}$ & $\epsilon^{1}$ & $\epsilon^{0}$ & $\epsilon^{1/2}$ & $\epsilon^{-a+1}$ & $\epsilon^{0}$ & $\epsilon^{-a+2}$
\\
(\ref{eq:sol2}) & $\epsilon^{1/2}$ & $\epsilon^{0}$ & $\epsilon^{1/2}$ & $\epsilon^{0}$ & $\epsilon^{1/4}$ & $\epsilon^{-a-(b-1)/2}$ ($\epsilon^{-a+1}$) & $\epsilon^{0}$ & $\epsilon^{-a-b/2+1}$ ($\epsilon^{-a+3/2}$)
\\
(\ref{eq:sol3}) & $\epsilon^{1}$ & $\epsilon^{0}$ & $\epsilon^{0}$ & $\epsilon^{0}$ & $\epsilon^{1}$ & $\epsilon^{-a-b}$ & $\epsilon^{-1}$ & $\epsilon^{-a-b+3}$
\\
(\ref{eq:sol4}) & $\epsilon^{1}$ & $\epsilon^{0}$ & $\epsilon^{0}$ & $\epsilon^{0}$ & $\epsilon^{1}$ & $\epsilon^{-a-b}$ & $\epsilon^{-1}$ & $\epsilon^{-a-b+3}$
\\
(\ref{eq:sol5}) & $\epsilon^{1/2}$ & $\epsilon^{0}$ & $\epsilon^{1/2}$ & $\epsilon^{0}$ & $\epsilon^{1/4}$ & $\epsilon^{-a-(b-1)/2}$ ($\epsilon^{-a+1}$) & $\epsilon^{0}$ & $\epsilon^{-a-b/2+1}$  ($\epsilon^{-a+3/2}$)
\\
(\ref{eq:sol6}) & $\epsilon^{0}$ & $\epsilon^{0}$ & $\epsilon^{1}$ & $\epsilon^{0}$ & $\epsilon^{1/2}$ & $\epsilon^{-a+1}$ & $\epsilon^{0}$ & $\epsilon^{-a+2}$
\end{tabular}
\end{center}
\end{table*}
Assuming the power-law spectrum of the waveaction, $n(\bm{k}, m) = |\bm{k}|^{-a} |m|^{-b}$,
we make Taylor expansion for the integrand of the kinetic equation (\ref{eq:keap})
as powers of $\epsilon$, that is $x$ and $y$.
Then, we get Table~\ref{table:zero} which shows
the leading orders of the each terms according to the asymptotics.
The leading order of the collision integral is given by ID when $-3<b<3$.
Therefore, we are going to show the procedure to get the leading order of ID (\ref{eq:sol2}) and (\ref{eq:sol5}) below.

As $\epsilon \to 0$,
$n_2 \to n$ for ID solutions.
Therefore,
the leading orders of
$f^0_{1,2} \sim n_1 (n_2 - n)$ and
$f^2_{0,1} \sim n_1 (n - n_2)$
is $O(\epsilon^{-a-(b-1)/2})$.
The order $O(\epsilon^{-a-b/2})$ is canceled as $\epsilon \to 0$.
This is called the {\em first cancellation\/}.
It must be noted that the leading order when $b=0$ is $1/2$ larger than that when $b \neq 0$
since $\partial n/\partial m = 0$.
The leading orders when $b=0$ are written in parentheses in Table~\ref{table:zero}.

The leading order of the integrand in Eq.~(\ref{eq:keap}) is written as
\begin{eqnarray}
&& 
T^0_{1,2} - T^1_{2,0} - T^2_{0,1} \propto k^{-2a + 3} m^{-2b+1}
\frac{x^{-a - (b+1)/2} y}{\sqrt{(x+y)(x-y)}}
\nonumber\\
&&
\times
\left(
-2 a y^2 \! - b \left((1-b) y (x+y) -2x(x-y)\right)  + b (b+1) xy
\right) .
\end{eqnarray}
Therefore,
the integrand has $O(\epsilon^{-a - (b - 5)/2})$.
The term which has $O(\epsilon^{-a -b/2 + 2})$ is canceled
since $T^2_{0,1} \to T^0_{1,2}$ (and $T^1_{2,0} \to 0$) as $\epsilon \to 0$.
This is the {\em second cancellation\/}.

Finally, we get the leading order of the kinetic equation after integration over $y$ from $-x$ to $x$:
\begin{eqnarray}
  \frac{\partial n_{\bm{p}}}{\partial t} &\propto&
- b (1-b) k^{4 - 2 a} m^{1 - 2b}
 \int_0  x^{-a - (b-5)/2} dx.
\label{ZeroContribution}
\end{eqnarray}
The integral has $O(\epsilon^{-a - (b - 7)/2})$.
Consequently integral converges if 
\begin{eqnarray*}
a + (b - 7)/2 <0 \;\;\mathrm{and}\;\; -3 < b < 3.
\end{eqnarray*}

The integral for the PR spectrum, which gives $O(\epsilon^{-1/4})$,
diverges as $k_1 \to 0$.
However, the integral for the GM spectrum converges
because $b=0$ and the next order is $O(1)$.

It should be noted that the leading order when $b=1$ is $1/2$ larger than that when $b \neq 0,1$
since a balance between first- and second-order derivative is made.
The leading orders when $b=1$ are $O(\epsilon^{-a+7/2})$.
It is also helpful to note that
$T^0_{1,2} - T^1_{2,0} - T^2_{1,0} = O(\epsilon^{-a+2})$ for ES
because of no second cancellation.
However, the collision integral has $O(\epsilon^{-a+5})$
because of symmetry of $y$.
Therefore, the integral which is dominated by ES converges
\begin{eqnarray*}
 a - 5 < 0 \;\;\mathrm{and}\;\; b < -3.
\end{eqnarray*}

Similarly, the integral which is dominated by PSI converges
\begin{eqnarray*}
 a + b - 5 < 0 \;\;\mathrm{and}\;\; b > 3.
\end{eqnarray*}

\subsection{Ultra-violet asymptotics}

\begin{table*}
  \caption{Asymptotics as $k_1 \to \infty$.
 PSI (\ref{eq:sol1}, \ref{eq:sol2}) gives $\epsilon^{a+b-3}$.
 ES (\ref{eq:sol3}, \ref{eq:sol5}) gives $\epsilon^{a-3}$.
 ID (\ref{eq:sol4}, \ref{eq:sol6}) gives $\epsilon^{a+b/2-4}$ ($\epsilon^{a-7/2}$).
The asymptotics for $b=0$ appear in parentheses.
}
  \label{table:infinity}
 \begin{center}
\begin{tabular}{ccccccccc}
 & $m_1$ & $m_2$ & $\omega_1$ & $\omega_2$ & $V^{\bm{p}_i}_{\bm{p}_j\bm{p}_k}$ & $f^{\bm{p}_i}_{\bm{p}_j\bm{p}_k}$ & $g^{i \prime}_{j, k}$ & $T^i_{j,k}$
\\
\hline
\hline
(\ref{eq:sol1}) & $\epsilon^{-1}$ & $\epsilon^{-1}$ & $\epsilon^{0}$ & $\epsilon^{0}$ & $\epsilon^{0}$ & $\epsilon^{a+b}$ & $\epsilon^{1}$ & $\epsilon^{a+b-2}$
\\
(\ref{eq:sol2}) & $\epsilon^{-1}$ & $\epsilon^{-1}$ & $\epsilon^{0}$ & $\epsilon^{0}$ & $\epsilon^{0}$ & $\epsilon^{a+b}$ & $\epsilon^{1}$ & $\epsilon^{a+b-2}$
\\
(\ref{eq:sol3}) & $\epsilon^{0}$ & $\epsilon^{0}$ & $\epsilon^{-1}$ & $\epsilon^{-1}$ & $\epsilon^{-1}$ & $\epsilon^{a}$ & $\epsilon^{-1}$ & $\epsilon^{a-2}$
\\
(\ref{eq:sol4}) & $\epsilon^{-1/2}$ & $\epsilon^{-1/2}$ & $\epsilon^{-1/2}$ & $\epsilon^{-1/2}$ & $\epsilon^{-1}$ & $\epsilon^{a+b/2+1}$ ($\epsilon^{a+3/2}$) & $\epsilon^{1/2}$ & $\epsilon^{a+b/2-3}$ ($\epsilon^{a-5/2}$)
\\
(\ref{eq:sol5}) & $\epsilon^{0}$ & $\epsilon^{0}$ & $\epsilon^{-1}$ & $\epsilon^{-1}$ & $\epsilon^{-1}$ & $\epsilon^{a}$ & $\epsilon^{-1}$ & $\epsilon^{a-2}$
\\
(\ref{eq:sol6}) & $\epsilon^{-1/2}$ & $\epsilon^{-1/2}$ & $\epsilon^{-1/2}$ & $\epsilon^{-1/2}$ & $\epsilon^{-1}$ & $\epsilon^{a+b/2+1}$ ($\epsilon^{a+3/2}$) & $\epsilon^{1/2}$ & $\epsilon^{a+b/2-3}$ ($\epsilon^{a-5/2}$)
\end{tabular}
\end{center}
\end{table*}

Next, we consider the limit $k_1 \to \infty$.
In this case, $k_2$ also approaches to infinity.
We employ the independent variables $x$ and $y$
as $k_1 = k/2 (1 + 1/x + y)$ and $k_2 = k/2 (1 + 1/x - y)$,
where $x = O(\epsilon)$ and $-1 < y < 1$.
Again, $m>0$ is assumed.

The leading orders are obtained by the similar manner used in the IR asymptotic
and are summarized in Table~\ref{table:infinity}.
The leading order of the integral is given by ID,
whose wavenumbers are given by Eqs.~(\ref{eq:sol4}, \ref{eq:sol6}),
when $-2 < b < 2$.
In this limit, no second cancellation is made.

As the result of the perturbation theory,
we get the leading order,
\begin{eqnarray}
 \frac{\partial n_{\bm{p}}}{\partial t} &\propto&
k^{4-2a} m^{1-2b}
b \int_0 x^{a+b/2-5} d x .
\label{InfinityContribution}
\end{eqnarray}
It has $O(\epsilon^{a+b/2-4})$.
Therefore, the integral converges if
\begin{eqnarray*}
a+b/2-4>0 \;\;\mathrm{and}\;\; -2<b<2 .
\end{eqnarray*}

The integral for the PR spectrum, which gives $O(\epsilon^{-1/4})$,
diverges as $k_1 \to \infty$.
and that for the Garrett--Munk spectrum, which gives $O(\epsilon^0)$,
converges owing that $b=0$.

Similarly,
\begin{eqnarray}
 \frac{\partial n_{\bm{p}}}{\partial t} &\propto&
- k^{4-2a} m^{1-2b}
b \int_0 x^{a-2}  d x
\label{InfinityContributionES}
\end{eqnarray}
for ES, which is dominant for $b>2$.
Consequently,
the integral converges also if
\begin{eqnarray*}
 a-3 > 0 \;\;\mathrm{and}\;\; b>2.
\end{eqnarray*}

In the same manner,
the convergent domain of the integral for PSI is given by
\begin{eqnarray*}
 a + b - 3 > 0 \;\;\mathrm{and}\;\; b < -2. 
\end{eqnarray*}

\end{appendix}

\begin{appendix}[B]
\section*{Frequency--vertical-wavenumber and horizontal--vertical-wavenumber
spectrum}
\label{OmegaMKMtransform}

The theoretical work presented below addresses the asymptotic power
laws of a three-dimensional action spectrum.  In order to connect with
that work, note that a horizontally isotropic power-law form of the
three-dimensional wave action $n(\bm{k},m)$ is given by
Eq.~(\ref{PowerLawSpectrum}).

The corresponding vertical wavenumber-frequency spectrum of energy
is obtained by transforming $ n_{\bm{k},m}$ from wavenumber space
$(\bm{k},m)$ to the vertical wavenumber-frequency space $(\omega,m)$ and
multiplying by frequency.  In the high-frequency large-wavenumber
limit,
$$ E(m,\omega)\propto \omega^{2-a} m^{2-a-b} ~~ .  $$

The total energy density of the wave field is then $$ E=\int \omega(\bm{k},m) n(\bm{k},m)\  d\bm{k} d m  = \int  E(\omega,m) \ d\omega d m  \label{TotalEnergy}.  $$
Thus, we also it convenient to work with the wave action spectrum
expressed as a function of $\omega$ and $m$
Therefore we also introduced (\ref{PowerLawSpectrumFrequencyVertical}).
The relation between $a$, $b$ and $\widetilde{a}$, $\widetilde{b}$ reads:
$$\widetilde{a}=-a, \quad \widetilde{b}=-a-b.$$

\end{appendix}

\begin{appendix}[C]
\section*{Asymptotic expansion for small $f$ values.\label{Appendixf}}
In this section we perform the small $f$ calculations of Section~\ref{CorRegularization}.
We start from the kinetic equation written as Eq.~(\ref{KE4}). There 
we  change variables in the first line of 
Eq.~(\ref{KE4}) as
$$m_1 = E_3(\omega_1)+ \mu(E_1(\omega_1) - E_3(\omega_1) ), $$
and in the second line of Eq.~(\ref{KE4}) as
$$m_1=E_3(\omega_1)+\mu (E_2({\omega_1}) - E_3({\omega_1})).$$ 
Then the  Eq.~(\ref{KE4}) becomes the following form:
\begin{eqnarray}
\frac{\partial}{\partial t} n(k(\omega,m),m) &=&\frac{2}{k}
  \int _f^{f+\omega_{\mathrm{s}}} d \omega_1 \int_0^1 d\mu {\cal P}_1
- {\frac{2}{k}} \int_{\omega-f-\omega_{\mathrm{s}}}^{\omega-f} d \omega_1
\int_0^1 d \mu {\cal P}_2
.
\label{KE5}
\end{eqnarray}
Here we introduced integrands ${\cal P}_1$ and ${\cal P}_2$ to be
\begin{eqnarray}
{\cal P}_1 = 
J\frac{|V^0_{1,2}|^2}{S^0_{1,2}} n_1 (n_2 - n) 
(E_1(\omega_1) - E_3(\omega_1)), \nonumber \\
{\cal P}_2 = 
J\frac{|V^0_{1,2}|^2}{S^0_{1,2}} n_2 (n - n_1) 
(E_2(\omega_1)-E_3(\omega_1)).
\label{IntegrandParts}
\end{eqnarray}

Before proceeding, note the following symmetries:
$$E_1(\omega_1 = \omega -  \omega_1^{\prime}) = m - E_2(\omega_1^{\prime}),$$
$$E_3(\omega_1 = \omega - \omega_1^{\prime}) = m  - E_3(\omega_1^{\prime}),$$
and
$$E_4(\omega_1 = \omega - \omega_1^{\prime}) = m  - E_4(\omega_1^{\prime}).$$
These symmetries explain why two disconnected regions on Figure
(\ref{KinematicBoxOM}) look like mirrored and shifted copies of each
other.  These symmetries further allows us simplification of
evaluation of ID contribution by noticing that contribution from ID1
is equal to contribution of ID3. 
To quantify the contribution of near-inertial waves to a $(\omega,m)$ mode, we write
$$\epsilon \sim f \ll \omega=1. $$
Subsequently, in the domain (a) we  write
$$\omega_1 = f + \epsilon,$$
in ${\cal P}_1$, and 
$$\omega_1 = \omega+f+\epsilon$$
in ${\cal P}_2$.
Furthermore, we expand ${\cal P}_1$ and ${\cal P}_2$ in powers of
$\epsilon$ and $f$ without making any assumptions of the relative
smallness of $f$ and $\epsilon$.
We use the facts that
$$
m>0, \quad \epsilon >0, \quad f>0, \quad 0<\mu<1.
$$
Define 
$${\cal P}_1 = P_1+P_2,$$
and 
$${\cal P}_2 = P_3+P_4.$$

This allows us to expand $P_1$, $P_2$, $P_3$ and $P_4$ in powers of $f$ and $\epsilon$.
We perform these calculations analytically on  
Mathematica software.

Mathematica was then able to 
perform the integrals of ${\cal P}_1$ and ${\cal P}_2$ over $\mu$
from $0$ to $1$ in Eq.~(\ref{KE5}) analytically. The result is given by
Eq.~(\ref{KE6}).

\end{appendix}

\bibliographystyle{ametsoc}

\end{document}